\documentclass[usenatbib,usenatbib]{mnras}
\usepackage{newtxtext,newtxmath}
\usepackage{todonotes}
\usepackage{times}
\usepackage{bm}
\usepackage{graphicx} 
\usepackage{bmpsize}
\usepackage{epstopdf}
\usepackage{dblfloatfix} 
\graphicspath{{./image/}}
\usepackage[normalem]{ulem}
\usepackage[T1]{fontenc}
\usepackage{makecell}
\usepackage{multirow}

\usepackage{comment}

\newcommand{\HI}{\rm H~{\sc i }}
\newcommand{\HII}{\rm H~{\sc ii }}

\newcommand{\TB}{\delta T_{\rm b}}
\newcommand{\MSUN}{{\rm M}_{\odot}}

\newcommand{\XHI}{x_{\rm HI}}
\newcommand{\XHII}{x_{\rm HII}}
\newcommand{\TS}{T_{\rm S}}
\newcommand{\TK}{T_{\rm K}}
\newcommand{\TCMB}{T_{\gamma}}
\newcommand{\TEFF}{T_{\gamma, \rm eff}}
\newcommand{\lya}{\rm {Ly{\alpha}}}
\newcommand{\OmegaB}{\Omega_{\rm B}}
\newcommand{\Omegam}{\Omega_{\rm m}}

\newcommand{\CommentOut}[1]{}

\title[IGM during the EoR]{Constraining the state of the intergalactic medium during the Epoch of Reionization using MWA 21-cm signal observations}

\author[Ghara et al.]{Raghunath Ghara$^{1, 2}$\thanks{E-mail: ghara.raghunath@gmail.com}, Sambit K. Giri$^{3}$\thanks{E-mail: sambit.giri@gmail.com}, Benedetta Ciardi$^{4}$, Garrelt Mellema$^{5}$,  and  \newauthor Saleem Zaroubi$^{1,6,2}$
\\
$^{1}$ARCO (Astrophysics Research Center), Department of Natural Sciences, The Open University of Israel, 1 University Road, PO Box 808, Ra'anana 4353701, Israel \\
$^{2}$Department of Physics, Technion, Haifa 32000, Israel\\
$^{3}$Institute for Computational Science, University of Zurich, Winterthurerstrasse 190, 8057 Zurich, Switzerland \\
$^{4}$Max-Planck Institute for Astrophysics, Karl-Schwarzschild-Stra{\ss}e 1, 85748 Garching, Germany\\
$^{5}$The Oskar Klein Centre, Department of Astronomy, Stockholm University, AlbaNova, SE-10691 Stockholm, Sweden\\
$^{6}$Kapteyn Astronomical Institute, University of Groningen, PO Box 800, 9700AV Groningen, the Netherlands\\
}

\date{Accepted XXX. Received YYY; in original form ZZZ}

\pubyear{}

\begin{document}
\label{firstpage}
\pagerange{\pageref{firstpage}--\pageref{lastpage}}
\maketitle
\begin{abstract}
The Murchison Widefield Array (MWA) team has derived new upper limits on the spherically averaged power spectrum of the 21-cm signal at six redshifts in the range $z \approx 6.5-8.7$. We use these upper limits and a Bayesian inference framework to derive constraints on the ionization and thermal state of the intergalactic medium (IGM) as well as on the strength of a possible additional radio background. We do not find any constraints on the state of the IGM for $z\gtrsim 7.8$ if no additional radio background is present. In the presence of such a radio background, the 95 per cent credible intervals of the disfavoured models at redshift $\gtrsim 6.5 $ correspond to an IGM with a volume averaged fraction of ionized regions below 0.6 and an average gas temperature $\lesssim 10^3$ K. In these models, the heated regions are characterised by a temperature larger than that of the radio background, and by a distribution with characteristic size $\lesssim 10$ $h^{-1}$ Mpc and a full width at half maximum (FWHM) of $\lesssim 30$ $h^{-1}$ Mpc. Within the same credible interval limits, we exclude an additional radio background of at least $0.008\%$ of the CMB at 1.42 GHz. 

\end{abstract}

\begin{keywords}
radiative transfer - galaxies: formation - intergalactic medium - cosmology: theory - dark ages, reionization, first stars - X-rays: galaxies
\end{keywords}


\section{INTRODUCTION}
\label{sec:intro}

During the Epoch of Reionization (EoR) ultra-violet (UV) radiation from the earliest generations of galaxies ionized the primordial neutral hydrogen (\HI) in the intergalactic medium (IGM). This global phase transition permanently changed the state of the IGM and had far reaching consequences for the subsequent formation and evolution of galaxies. Observations of the absorption spectra of quasars at redshift $z\gtrsim6$  
\citep[e.g.][]{Fan06b,2018Natur.553..473B, 2018ApJ...864..142D, 2017MNRAS.466.4239G,2019MNRAS.484.5094G,  2020ApJ...896...23W, 2020ApJ...904...26Y} and the measurement of the Thomson scattering optical depth from observations of the Cosmic Microwave Background (CMB) \citep{2020A&A...641A...6P} suggest that this transition took place sometime between redshift $\approx$ 6 and 10  \citep[e.g,][]{Mitra15}. However, many  of its details, such as the exact timing of the process, the ionization and thermal state of the IGM, the morphology of the ionized regions, remain poorly understood \citep[see e.g.][for reviews]{Pritchard12, 2013ASSL..396...45Z}. 

The 21-cm signal produced by \HI in the IGM is the most direct probe of the reionization process as it depends directly on the \HI density. It is therefore expected to answer many of the outstanding questions, including the nature of the sources responsible for reionization. Thus, substantial efforts have been made to measure this signal. Observations of the 21-cm signal fall into two categories. The first type uses large interferometric radio telescopes to measure the statistical fluctuations of the signal. Examples of such telescopes are the Low Frequency Array (LOFAR)\footnote{\url{http://www.lofar.org/}} \citep{vanHaarlem2013LOFAR:ARray,2017ApJ...838...65P}, the Precision Array for Probing the Epoch of Reionization (PAPER)\footnote{\url{http://eor.berkeley.edu/}} \citep{parsons13, 2019ApJ...883..133K}, the Murchison Widefield Array (MWA)\footnote{\url{http://www.mwatelescope.org/}} \citep[e.g.][]{tingay13,Wayth2018mwa}, the New extension in Nan\c{c}ay Upgrading loFAR (NenuFAR)\footnote{\url{ https://nenufar.obs-nancay.fr/}} \citep{2012sf2a.conf..687Z}, the Upgraded Giant Metrewave Radio Telescope (uGMRT) \citep{2017CSci..113..707G}, the Owens Valley Radio Observatory Long Wavelength Array (OVRO-LWA)\footnote{\url{http://www.tauceti.caltech.edu/LWA/}} 
and the Hydrogen Epoch of Reionization Array (HERA)\footnote{\url{https://reionization.org/}} \citep{2017PASP..129d5001D}. A more sensitive radio interferometer, the upcoming Square Kilometre Array (SKA)\footnote{\url{http://www.skatelescope.org/}}, should be able to also produce tomographic images of the distribution of the signal on the sky \citep{2015aska.confE..10M, ghara16, giri2017bubble, 2018arXiv180106550G, Giri1289039, 2020MNRAS.496..739G,giri2020BettiNumbers, 2021arXiv210103962K}. 
The second type of observations use single antennae and attempt to measure the sky-averaged 21-cm signal and its evolution with time. Examples are EDGES \citep{2010Natur.468..796B}, EDGES2 \citep{monsalve2017,EDGES2018}, SARAS \citep{2015ApJ...801..138P}, SARAS2 \citep{singh2017}, SARAS3 \citep{saras3}, PRIZM \citep{2019JAI.....850004P}, BigHorns \citep{2015PASA...32....4S}, SciHi  \citep{2014ApJ...782L...9V} and LEDA \citep{price2018}.

One of the main challenges of these experiments is to remove the contribution of the galactic and extra-galactic foregrounds, which is stronger than the expected 21-cm signal by several orders of magnitude. Their smooth frequency dependence should allow a separation from the fluctuating 21-cm signal, allowing them to be either subtracted \citep{2009MNRAS.397.1138H, 2015MNRAS.447.1973B, 2016MNRAS.458.2928C, 2018MNRAS.478.3640M, 2021MNRAS.500.2264H}, suppressed \citep{kanan2007MNRAS.382..809D, 2012MNRAS.426.3178M, ghara15c} or avoided \citep{2010ApJ...724..526D, 2014PhRvD..90b3019L}. Moreover, an exquisite calibration of the system  is required to minimize the artefacts from strong sources  \citep{2016MNRAS.461.3135B, 2017ApJ...838...65P}, and a long observation time is necessary to reduce the instrumental noise. Further calibration challenges may be posed by the hardware components of the telescope \citep[see e.g.,][]{2019ApJ...884..105K} and by the temporally and spatially varying ionosphere \citep[see e.g.,][]{2016RaSc...51..927M}, to mention a few.

So far, no undisputed detection of the 21-cm signal from the EoR has been achieved. \citet{EDGES2018} analysed observations with the EDGES2 low-band antenna  and claimed a detection of the averaged  brightness temperature of the 21-cm signal ($\overline{\TB}$) at various redshifts. The redshift evolution of $\overline{\TB}$ has a minimum  at $z\approx 17$ with an amplitude much stronger than predicted by standard theories. Therefore the validity of these claims have been debated \citep[e.g.\ in][] {2018Natur.564E..32H, 2018ApJ...858L..10D, 2019ApJ...880...26S, 2019ApJ...874..153B}, and unconventional physical processes such as an unknown cooling mechanism \citep[see e.g.,][]{2014PhRvD..90h3522T, 2018Natur.555...71B, 2018PhRvL.121a1101F, 2018arXiv180210094M, 2020MNRAS.492..634G} or the presence of an additional radio background on top of the CMB \citep[][]{2018ApJ...858L..17F, 2018ApJ...868...63E, 2019MNRAS.486.1763F, Mebane2020} have been invoked to explain these results.

While none of the other attempts have to date claimed a detection of the signal, several of them have provided upper limits.  A 2$\sigma$ value of $\Delta^2(k) < (248)^2~ \rm mK^2$ for a $k$-scale $0.5 ~h ~\rm Mpc^{-1}$  was the very first upper limit on the power spectrum\footnote{Here and for the rest of this paper we use the so-called dimensionless form of the power spectrum, $\Delta^2(k)$, which is related to the standard spherically averaged power spectrum $P(k)$ through $\Delta^2(k) = k^3 P(k)/2\pi^2$.} of the 21-cm signal from redshift $8.6$, obtained from observations with the GMRT \citep{paciga13}.  \citet{2019ApJ...884....1B}, \citet{2018ApJ...868...26C}, and \citet{2019ApJ...883..133K} provided upper limits using the observations with MWA and PAPER. \citet{2017ApJ...838...65P} published the first upper limit on the power spectrum of the signal from redshift between 9.6 and 10.6 using LOFAR observations. Later, \citet{2019MNRAS.488.4271G,2020MNRAS.499.4158G} and \citet{2019AJ....158...84E} probed a higher redshift range and also provided upper limits on the power spectrum using observations with the LOFAR-Low Band Antenna array and the OVRO-LWA, respectively. Recently, \citet{2020MNRAS.493.1662M} analysed 10 nights of LOFAR observations of the North Celestial Pole and obtained a 2$\sigma$ upper limit on the 21-cm signal power spectrum at $z\approx 9.1$ of $\Delta^2(k=0.075 ~h ~\rm Mpc^{-1})=(73)^2 \rm ~mK^2$.

The focus of this paper is on the upper limits in the redshift range 6.5 -- 8.7 published by the MWA team \citep{2020MNRAS.493.4711T}. Their best 2$\sigma$ upper limit at redshift 6.5 is $\Delta^2(k = 0.14 ~h ~\rm Mpc^{-1})\approx (43)^2 \rm ~mK^2$, derived from 110 hours of integration with the MWA high band on the EoR0 field. The upper limit values increase with redshift. At $z=8.7$, the upper limit of  $\Delta^2(k = 0.14 ~h ~\rm Mpc^{-1})\approx (250)^2 \rm ~mK^2$ comes from 51 hours of observations of the EoR2 field.

The emitted 21-cm signal from \HI depends on the number density of the sources present at those epochs, as well as on the emissivity of UV, X-ray, radio and $\lya$ photons from those sources \citep[see e.g.,][]{2019MNRAS.tmp.1183R, 2020MNRAS.499.5993R}. Due to the complex dependence of the signal on the sources, it is not straightforward to extract the astrophysical as well as the cosmological information from 21-cm signal observations, and exploration of many theoretical models of the signal is required to interpret the results. Often, the signal prediction algorithms are used in a Bayesian inference framework to explore the parameter space of the reionization models \citep[e.g.][]{Greig201521CMMC:Signal,2017MNRAS.472.2651G,Park2019InferringSignal, 2020MNRAS.495.4845C, 2020MNRAS.493.4728G, 2020MNRAS.498.4178M, 2020arXiv200802639G, 2020arXiv200603203G}.

In this paper, we consider the new upper limits on the 21-cm signal power spectrum from MWA observations \citep{2020MNRAS.493.4711T} in the redshift range spanning from 6.5 to 8.7 to constrain the IGM conditions and explore scenarios for the EoR that are disfavoured by such limits. Our framework is based on simulations of the power spectrum and IGM parameters and a Bayesian inference framework. Just as in our work on the LOFAR upper limits for $z\approx 9.1$ \citep{2020MNRAS.493.4728G}, our main focus is on deriving constraints on the condition of the IGM. In addition, we also aim to put constraints on the additional radio background which has been proposed to explain the EDGES2 results. This work differs from the interpretation of the MWA upper limits by \citet{2020arXiv200802639G}, who instead focused on the properties of the sources and did not consider any additional radio background component to the CMB.

This paper is structured as follows. In Section \ref{sec:method}, we describe the basic methodology of our framework.  We present our results in Section \ref{sec:results}. We discuss the outcome of this study from the point of view of other observations in Section \ref{sec:discussion}, before concluding in Section \ref{sec:con}.
The cosmological parameters used throughout this study are the same of the $N$-body simulations employed here, i.e.  $\Omegam=0.27$, $\Omega_\Lambda=0.73$, $\OmegaB=0.044$, $h=0.7$  \citep[Wilkinson ~Microwave ~Anisotropy ~Probe (WMAP);][]{2013ApJS..208...19H}.

\section{Framework}
\label{sec:method}

Here we describe the framework employed to constrain the IGM properties using the observational upper limits from the MWA.


\begin{table*}
\centering
\caption[]{Overview of the simulation parameters used in {\sc grizzly} and their explored ranges.}
\small
\tabcolsep 2pt
\renewcommand\arraystretch{1.5}
   \begin{tabular}{c c c c c}
\hline
\hline
Simulation Parameters & Description & Explored range  	 \\
\hline
\hline
$\zeta$ & Ionization efficiency   & [$10^{-3}, 10^{3}$]   &   	\\
$M_{\rm min}$ & Minimum mass of the UV emitting halos   & [$10^{9}  ~\MSUN, 10^{12} ~\MSUN$]        & 	\\
$M_{\rm min,X}$ & Minimum mass of the X-ray emitting halos   & [$10^{9} ~\MSUN, 10^{12} ~\MSUN$]   &  	\\
$f_X$ & X-ray heating efficiency   & [$10^{-3}$, $10^2$]   & 	\\
$A_r$ & Efficiency of the excess radio background & [0, 416]  & \\
\hline
\hline
\end{tabular}
\label{tab_source_param}
\end{table*}

\subsection{Modelling the 21-cm signal using {\sc grizzly}}
\label{sec:grizzly}

Our framework uses the {\sc grizzly} code \citep{ghara15a, ghara18} to generate 21-cm signal maps for the redshift range $z\approx 6.5-8.7$ for different combinations of astrophysical parameters. The code requires gridded versions of cosmological density fields and halo catalogues as input. We use density fields of length 500~$h^{-1}$~ comoving megaparsec (Mpc) smoothed onto $300^3$ grids  \citep[see e.g,][]{2019MNRAS.489.1590G, 2019JCAP...02..058G, 2021MNRAS.502.3800K}, which were retrieved from the  results of the PRACE\footnote{Partnership for Advanced Computing in Europe: \url{http://www.prace-ri.eu/}} project PRACE4LOFAR. This box size corresponds to a field-of-view of $4.3^{\circ}\times 4.3^{\circ}$ at redshift $\approx$ 9. The minimum mass of the dark matter halos as identified on the fly with a spherical overdensity halo finder \citep{Watson2013TheAges} is $\approx 10^9$~M$_\odot$. These are the same  cosmological $N$-body simulation previously used in \citet{2020MNRAS.493.4728G}.

The {\sc grizzly} code is an independent implementation of the previously developed {\sc bears} algorithm  \citep{Thom08,Thom09,Thom11,2018NewA...64....9K}. It is based on a one-dimensional radiative transfer scheme which approximates the transfer of photons (UV, X-rays and $\lya$) by assuming that the effect from individual sources is isotropic. We refer the reader to the original papers \citep{ghara15a, ghara18} for a more detailed and complete description of the algorithm.

Given a set of astrophysical source parameters, {\sc grizzly} generates ionization and gas temperature maps at different redshifts which are then used to produce differential brightness temperature ($\TB$) maps. The $\TB$ of the 21-cm signal can be expressed as \citep[see e.g,][]{madau1997, Furlanetto2006}, 
\begin{eqnarray}
 \TB (\mathbf{x}, z) \!  & = & \! 27 ~ x_{\rm HI} (\mathbf{x}, z) [1+\delta_{\rm B}(\mathbf{x}, z)] \left(\frac{\OmegaB h^2}{0.023}\right) \nonumber\\
&\times& \!\left(\frac{0.15}{\Omegam h^2}\frac{1+z}{10}\right)^{1/2}  \left(1-\frac{\TEFF}{\TS(\mathbf{x}, z)} \right)\,\rm{mK},
\nonumber \\
\label{eq:brightnessT}
\end{eqnarray}
where  $x_{\rm HI}$ and $\delta_{\rm B}$ denote the neutral hydrogen fraction and baryonic density contrast respectively, each at position $\mathbf{x}$ and redshift $z$. The quantity $\TEFF$ is the temperature of the radio background, which reduces to the usual CMB temperature, $\TCMB(z)$ = $2.725~(1+z)$ K, in the absence of additional contributions. We assume that the spin temperature of hydrogen in the IGM,  $\TS$, is equal to the gas temperature. This condition occurs in the presence of a strong $\lya$ background which is expected at the redshifts of interest \citep[e.g.][]{barkana05b,Pritchard12,2020arXiv201112308S}.

Afterwards we calculate the dimensionless power spectrum of $\TB$, denoted as $\Delta^2(k)$. Note that we have ignored the presence of redshift space distortions (RSD) while evaluating the power spectrum for different model parameters, as their impact remains rather small during the epoch of our interest\footnote{The effects of RSD can enhance the large-scale power spectrum by a factor of $\approx 2$ in case it is driven by density fluctuations. In our case, we underestimate the power spectra values for those scenarios. }, when ionization fluctuations dominate the power spectrum of $\TB$ \citep{Jensen13, 2016JApA...37...32M, ghara15a, ghara15b, 2020arXiv201103558R}.

\subsection{Simulation parameters in {\sc grizzly}}
\label{sec:source_param}

{\sc grizzly} uses the following set of parameters to generate the maps of ionization fraction and gas temperature in the IGM. These are later used to generate $\TB$ maps and from these the power spectra. We describe the parameters below and summarise the range of values explored in Table \ref{tab_source_param}.

\begin{itemize}
    \item Ionization efficiency ($\zeta$): This scales the rate of emission of ionizing photons per unit stellar mass from a halo $\dot N_i$ as $\dot N_i=\zeta\times 2.85\times 10^{45}  ~{\rm s^{-1}} ~\MSUN^{-1}$. $\zeta=1$ corresponds to a model galaxy spectrum which has been produced with the publicly available population synthesis code {\sc pegase2} \footnote{\tt \url{http://www2.iap.fr/pegase/}} \citep{Fioc97} and employed in the 1D radiative transfer. We assume that the stellar mass $M_\star$ of a halo is related to the dark matter halo mass  $M_{\rm halo}$  as $M_\star = f_\star ~ \frac{\OmegaB}{\Omegam} ~ M_{\rm halo}$, where  the star formation efficiency $f_\star$ is fixed at 0.02 \citep{2015ApJ...799...32B, 2016MNRAS.460..417S}. Note that the effect of $\zeta$ is degenerate with that of $f_\star$, the emission rate of ionizing photons from the sources, and their escape fraction into the IGM. Here we vary Log$_{10}$($\zeta$) between [-3, 3].

    \item Minimum mass of the UV emitting halos ($M_{\rm min}$): We assume no emission of ionizing photons from halo with mass less than $M_{\rm min}$. This is to incorporate the fact that various mechanism such as the radiative and mechanical feedback can reduce the star formation efficiency significantly \citep[see e.g.,][]{2013MNRAS.428..154H, 2018MNRAS.480.1740D}. In general, one expects that the radiative feedback suppresses star formation in halos with mass $\lesssim 10^9 ~\MSUN$  \citep[see e.g.,][]{2014MNRAS.442.2560W, Dixon2016TheReionization}.  In our study the lowest value for $M_{\rm min}$ is $10^9$~$\MSUN$. This is limited by the mass resolution of our N-body simulation (see Sect.~\ref{sec:grizzly}). We vary  Log$_{10}$($\frac{M_{\rm min}}{\MSUN}$) between [9, 12].

    \item Minimum mass of X-ray emitting halo ($M_{\rm min, X}$): In addition to the UV emitting sources, {\sc grizzly} also includes heating and ionization from X-ray emitting sources, such as quasars and high-mass X-ray binaries, as not necessarily all halos hosting stellar sources also host X-ray sources. We assume no emission of X-ray photons from halos with mass less than $M_{\rm min, X}$, while all other halos host X-ray emitting sources. We vary  Log$_{10}$($\frac{M_{\rm min, X}}{\MSUN}$) between [9, 12].

    \item X-ray heating efficiency ($f_X$): This scales the emission rate of X-ray photons per unit stellar mass from a halo as $\dot N_X = f_X \times  10^{42} ~\rm s^{-1} ~\MSUN^{-1}$. We model the spectrum of an X-ray source as a power-law of the energy $E$, i.e. $I_X(E) \propto E^{-\alpha}$, where we fix the spectral index $\alpha$ to 1.2. The value of $\dot N_X$ for $f_X=1$ is consistent with the measurements of high-mass X-ray binaries in local star forming galaxies in the 0.5-8 keV band \citep{2012MNRAS.419.2095M, 2019MNRAS.487.2785I}.  We assume that the X-ray band spans from 100 eV to 10 keV. We vary Log$_{10}$($f_X$) between [-3, 2].

    \item Radio background efficiency ($A_r$): This parameter quantifies the contribution of a potential radio background existing in addition to the CMB. For the LOFAR upper limits this parameter was not considered in \citet{2020MNRAS.493.4728G} but was instead studied in a separate paper by \citet{2020MNRAS.498.4178M}. We model the effective brightness temperature of the radio background as \citep{2019MNRAS.486.1763F,2020MNRAS.498.4178M}, 
\begin{equation}
\TEFF = T_{\gamma} \left[1+A_r \left(\frac{\nu_{\rm obs}}{78 ~\rm MHz} \right)^{-2.6} \right], 
\label{eq:trad}
\end{equation}
where $\nu_{\rm obs}$ is the frequency of observation.
$\TEFF$ can differ from $\TCMB$ in presence of a radio background of any astrophysical or cosmological origin. Its existence is motivated by the evidence of an excess radio background above the CMB towards the Rayleigh-Jeans part of the spectrum, as detected by ARCADE2 \citep{2011ApJ...734....5F} and LWA1 \citep{2018ApJ...858L...9D}. The LWA1 measurement of the excess radio background at frequency 40-80 MHz fits well with a power-law with spectral index -2.58$\pm$0.05 with a temperature of $603^{+102}_{-92}$ mK at 1.42 GHz.  Varying $\TEFF$ changes the brightness temperature maps although the ionization and gas temperature maps from {\sc grizzly} remain unchanged. We vary $A_r$ from 0 to 416, where the lower bound stands for no excess radio background and the upper bound represents the upper limits of 0.603~K as measured by LWA1. Note that the radio background used in this study is uniform, while in general an inhomogeneous radio background may enhance the power spectrum even more than what we estimate here \citep[][]{Reis2020Highredshift}. A study with inhomogeneous radio background is beyond the scope of this paper.  
\end{itemize}

\begin{table}
\centering
\caption[]{An overview of the IGM parameters considered in this paper. Note that the `heated region' denotes the regions in the IGM with gas temperature larger than $\TEFF$.  }
\small
\tabcolsep 8pt
\renewcommand\arraystretch{1.5}
   \begin{tabular}{c c c  }
\hline
\hline
IGM Parameters & Description  &	 \\
\hline
\hline
$\overline{\XHII}$ & Volume averaged ionized fraction       &	\\
$\overline{\TK}$ (K) & \makecell{Volume averaged gas temperature in \\ the partially ionized IGM with $\XHII<0.5$}         &	\\
$\overline{\TB}$ (mK) & \makecell{Mass averaged \\ differential brightness temperature}         &	\\
$f_{\rm heat}$ & \makecell{Volume fraction of regions with \\ temperature larger than $\TEFF$}   &\\
$R^{\rm heat}_{\rm peak}$ (Mpc/h) & \makecell{Size at which the PDF of the size \\ distribution of the heated regions peaks}         & 	\\
$\Delta R^{\rm heat}_{\rm FWHM}$ (Mpc/h) & \makecell{FWHM of the PDF of the size \\ distribution of the heated regions}       &	\\
\hline
\hline
\end{tabular}
\label{tab_igm_param}
\end{table}

\subsection{Derived IGM parameters}
\label{sec_igm_param}
One should realise that the 21-cm signal observations do not contain any direct information about the properties of the sources at the redshift of interest. Instead, the 21-cm signal characterises the state of the IGM at those redshifts. 
Thus, the main focus of this study is on the IGM properties rather than on the source properties, i.e. we aim to constrain the physical properties of IGM, such as the ionization and thermal state, rather than the astrophysical simulation parameters used to simulate the brightness temperature maps. Below we list the IGM parameters that we consider here at each redshift (see also in Table \ref{tab_igm_param}):
\begin{itemize}
    \item $\overline{\XHII}$: Volume averaged ionized fraction of the IGM.
    \item $\overline{\TK}$ (K): Volume averaged gas temperature of the regions in the IGM with ionization fraction $\XHII< 0.5$.
    \item $\overline{\TB}$ (mK):  Mass averaged differential brightness temperature in the IGM.
    \item $f_{\rm heat}$: Volume fraction of the IGM with gas temperature $\TK>\TEFF$. 
    \item $R^{\rm heat}_{\rm peak}$ (cMpc): Size of the `heated regions', i.e. regions with $\TK>\TEFF$ at which the probability distribution function (PDF) of the sizes peaks. To determine the size distribution of the heated regions we use the mean free path (MFP) method, which is a Monte Carlo based approach   \citep{2007ApJ...669..663M,giri2017bubble}.
    \item $\Delta R^{\rm heat}_{\rm FWHM}$ (cMpc): Full width at half maximum (FWHM) of the PDF of the sizes of the heated regions.
\end{itemize}

Unlike in \citet{2020MNRAS.493.4728G}, here we have not considered separate IGM parameters for the \HII regions. In the absence of X-ray heating, the `heated regions' in the IGM are equivalent to the \HII regions, as the kinetic temperature of the gas within such regions is $\sim 10^4$K. Thus, for insignificant heating in the neutral IGM, $f_{\rm heat}$ is equivalent to $\overline{\XHII}$, while $R^{\rm heat}_{\rm peak}$ and $\Delta R^{\rm heat}_{\rm FWHM}$ characterise the size distribution of the \HII regions. Note that we estimate these quantities from {\sc grizzly} simulations at our simulation resolution.

\subsection{Bayesian inference framework}
\label{sec:like}

We generate $10^5$ power spectra and sets of IGM parameters at each redshift point using {\sc grizzly}, while employing the gridded sampling method to sample the 5D parameter space. We then use a linear interpolation scheme\footnote{Note that in our previous study \citep{2020MNRAS.493.4728G} we used an emulator based on the Gaussian Process Regression (GPR) algorithm. However, GPR does not scale well with the size of the training set \citep[e.g.][]{GPR2006}. As the parameter space used here has a higher dimension than the one explored in \citet{2020MNRAS.493.4728G}, a larger training set is needed. Tests showed that this would lead the emulator to achieve a size of several tens of gigabytes, slowing it down too much to be useful.}
on a regular grid to determine the power spectrum for a set of parameters $\theta$. To avoid extrapolation, the range of $\theta$ used is the same as the range of the gridded parameter space covered by the {\sc grizzly} simulations performed for this study. 
We use the regular grid interpolation method implemented in {\sc scipy} python package \citep{weiser1988note,Virtanen2020scipy}.
See \citet{2018ApJ...863...11M} for a similar Bayesian framework that is based on interpolation in a regular grid of 4D parameter space. This linear interpolator is used in the {\sc emcee} python MCMC module \citep{emcee2013paper} to explore the parameter space as described in section \ref{sec:source_param}. Note that we use only the smallest three $k$-bin values from \citet{2020MNRAS.493.4711T}, i.e. 0.142, 0.212 and 0.283 $h~ \rm Mpc^{-1}$, as the upper limits at other wavenumbers are too high to give any additional constraints.

\begin{table*}
\centering
\caption[]{The $\Delta^2_\mathrm{up} \equiv \Delta^2_\mathrm{21}+2\sigma_{21, \rm err}$ upper limits from MWA observations at different redshifts for different $k$-bins. See \citet{2020MNRAS.493.4711T} for details. We obtained $\Delta^2_\mathrm{21}$ and the corresponding error from private communication with the MWA team.}
\small
\tabcolsep 8pt
\renewcommand\arraystretch{1.5}
   \begin{tabular}{c c c c c c c c}
\hline
\hline
k ($h ~\rm Mpc^{-1}$) & \makecell{$\Delta^2_\mathrm{up}(k)$ (mK$^2$) \\ at $z$= 6.5} & \makecell{$\Delta^2_\mathrm{up}(k)$ (mK$^2$) \\ at $z$= 6.8}  & \makecell{$\Delta^2_\mathrm{up}(k)$ (mK$^2$) \\ at $z$= 7.1}  & \makecell{$\Delta^2_\mathrm{up}(k)$ (mK$^2$) \\ at $z$= 7.8} & \makecell{$\Delta^2_\mathrm{up}(k)$ (mK$^2$) \\ at $z$= 8.2} & \makecell{$\Delta^2_\mathrm{up}(k)$ (mK$^2$) \\ at $z$= 8.7} \\

\hline
\hline
0.142 & $43.1^2$  & $60.1^2$ & $77.7^2$ & $154.2^2$ & $167.7^2$ & $249.6^2$ &\\
0.212 & $70.2^2$ & $90.0^2$  & $117.4^2$ & $247.5^2$ & $430.3^2$ & $569.9^2$ &\\
0.283 & $93.3^2$ & $114.1^2$ & $152.3^2$ & $314.5^2$ & $422.2^2$ & $562.5^2$ &\\
0.354 & $209.5^2$ & $243.9^2$  & $281.5^2$ & $460.1^2$ & $540.9^2$ & $688.1^2$ &\\
0.425 & $183.5^2$ & $221.3^2$  & $263.3^2$ & $804.4^2$ & $772.8^2$ & $963.2^2$ &\\
0.495 & $125.5^2$ & $169.0^2$  & $231.9^2$ & $466.8^2$ & $1402.6^2$ & $1854.5^2$ &\\
0.566 & $210.1^2$ & $255.4^2$  & $310.9^2$ & $484.4^2$ & $1109.9^2$ & $1546.0^2$ &\\
0.637 & $214.1^2$ & $260.3^2$  & $333.8^2$ & $501.0^2$ & $739.1^2$ & $962.3^2$ &\\
0.708 & $384.6^2$ & $383.1^2$  & $437.9^2$ & $613.4^2$ & $781.1^2$ & $947.6^2$ &\\

\hline
\hline
\end{tabular}
\label{tab_obs}
\end{table*}

The MCMC analysis requires the calculation of a likelihood. As we are working with upper limits we cannot use the formal $\chi^2$ which only applies for actual measurements of a signal. Also, as the upper limits have quite high values, we want our framework to find the models which are excluded, rather than the ones which are allowed. Given a measured upper limit of the power spectrum at redshift $z$ as $\Delta^2_\mathrm{21} (k_i, z) \pm \Delta^2_\mathrm{21,err} (k_i, z)$, the likelihood of a set of parameters $\bm{\theta}$ to be excluded is defined as \citep[see][ for details]{2020MNRAS.493.4728G}
\begin{equation}
    \mathcal{L}_{\rm ex}(\theta, z) = 1- \prod_{i} \frac{1}{2}\left[ 1 + \mathrm{erf}\left(\frac{\Delta^2_\mathrm{21} (k_i, z)- \Delta^2_\mathrm{m}(k_i, \bm{\theta}, z)}{\sqrt{2} \sigma (k_i, z)}\right) \right],
    \label{equ:like}
\end{equation}
where $k_i$ denotes the 3 $k$-bins of the measured power spectrum, and $\Delta^2_\mathrm{m}(k, \bm{\theta}, z)$ the model power spectrum for a set of parameters $\theta$ for $k$-scale at redshift $z$. The quantity $\sigma$ accounts for errors from the measurement, as well as the modelling error, i.e. $\sigma^2 = \Delta^4_\mathrm{21,err} + \Delta^4_\mathrm{m,err}$. 
We adopt 30 per cent modelling error $\Delta^2_\mathrm{m,err} (k_i, z) = 0.3\times \Delta^2_\mathrm{m}(k_i, \bm{\theta}, z)$ to incorporate errors from the modelling with the code {\sc grizzly} and from the interpolation scheme. In \citet{ghara18} we performed a detailed comparison of {\sc grizzly} with the 3D radiative transfer scheme C$^2${\sc ray}, and found that the error from {\sc grizzly} is less than 10 per cent. We estimate the error due to the interpolation scheme on a test set of the power spectra and found it to be less than 30 per cent for the scales of interest. Note that the sample variances on the power spectrum at the scales of interest are negligible compared to the adopted modelling errors. 

\begin{figure*}
\begin{center}
\includegraphics[scale=0.4]{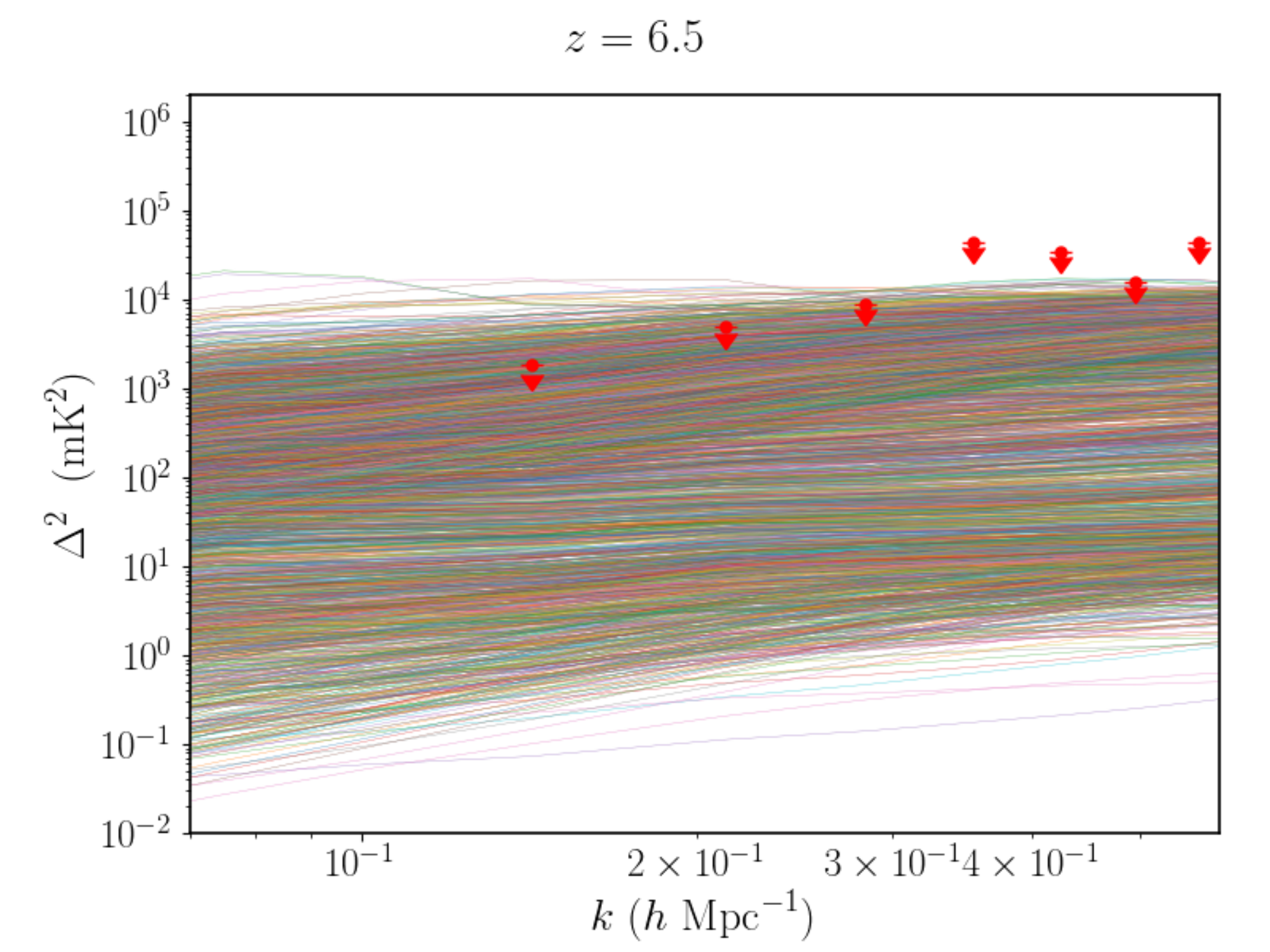}
\includegraphics[scale=0.4]{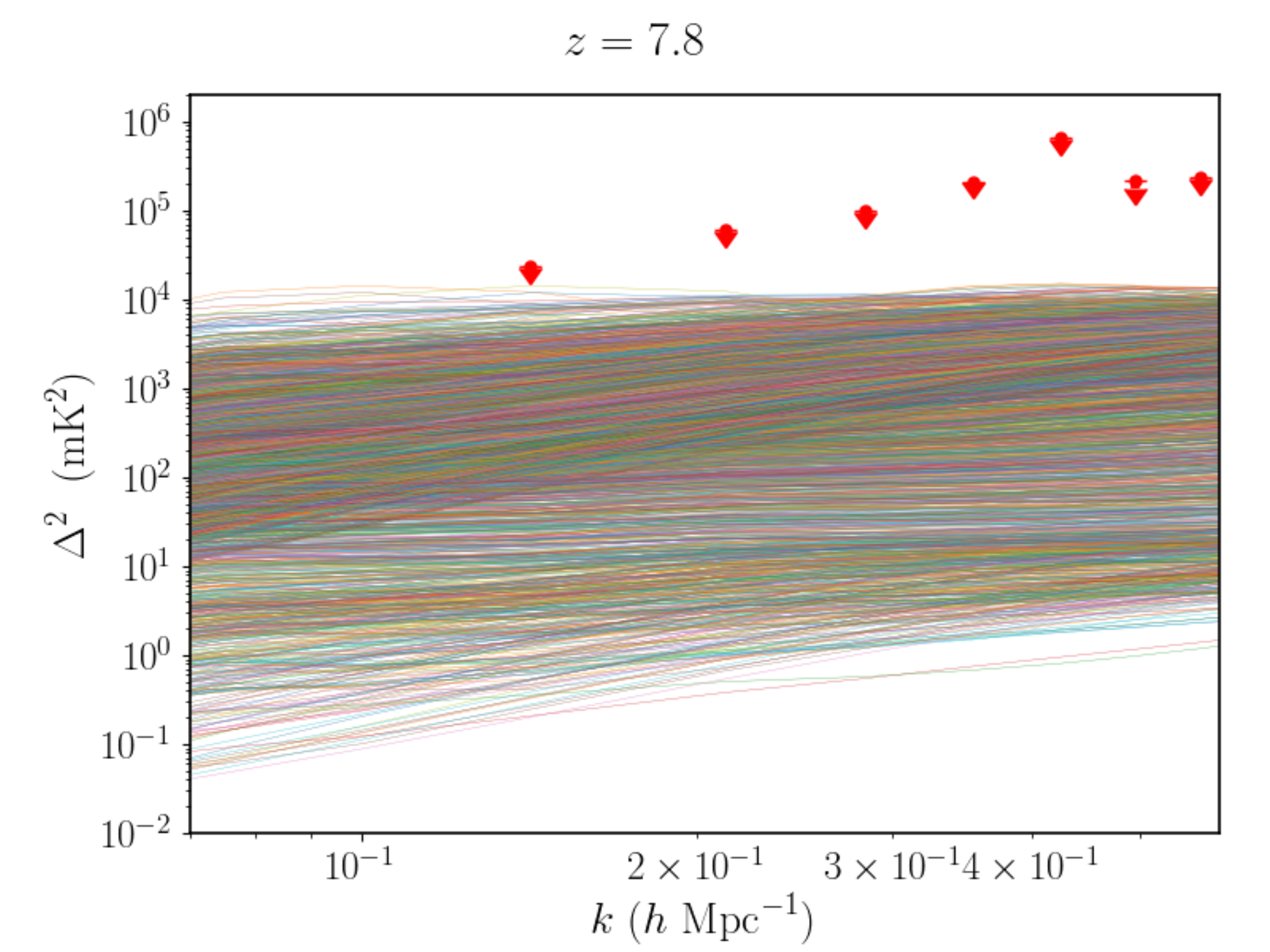}
    \caption{A subset of the power spectra of the 21-cm signal brightness temperature generated using {\sc grizzly} over the gridded parameter space. These correspond to $A_r = 0$ or, equivalently, $T_{\gamma, \rm eff}=2.725\times (1+z)$ K. Note that we have also varied $A_r$ to get the full power spectra set. Left and right panels correspond to $z\approx 6.5$ and $z\approx 7.8$, respectively. The red points refer to the recent MWA 2$\sigma$ upper limits from \citet{2020MNRAS.493.4711T}. }
   \label{image_psts_z}
\end{center}
\end{figure*}

Unlike \citet{2020arXiv200802639G}, we consider the upper limits at each redshift independently to obtain constraints on the simulation parameters and thereafter on the IGM parameters. One requires a background source model to derive constraints on the simulation parameters using multi-redshift constraints. Thus, the derived constraints on the IGM parameters at individual redshifts from such a combined upper limits analysis would be source model dependent. Moreover, those constraints would be biased as the likelihood in that case would be mostly dominated by the best upper limit at $z\approx 6.5$ (see Table \ref{tab_obs}).    Thus, to avoid dependence of the quantities of our interest on the background source model, we consider the upper limits at each redshift independently.

Once the MCMC chains that contain the lists of the simulation parameters used in the MCMC analysis is built, we generate the list of corresponding IGM parameters using the interpolation scheme.  Finally, these are used to determine the posterior distribution of the IGM parameters.

\section{Results}
\label{sec:results}
In this section, we apply the Bayesian inference framework described in the previous section to the measured upper limits from \citet{2020MNRAS.493.4711T}. These 2$\sigma$ upper limits are also shown in Table \ref{tab_obs}. Clearly, the upper limits are stronger at the larger (spatial) scales (lower `$k$' values) and lower redshifts.
Figure \ref{image_psts_z} shows the set of the power spectra generated with {\sc grizzly} for $A_r=0$ at redshift 6.5 (left panel) and 7.8 (right panel), respectively. The red points in the panels denote the measured MWA 2$\sigma$ upper limits on the power spectrum. Clearly, the power spectrum amplitude of a significant fraction of the models is above the upper limit at scales $\lesssim 0.2 ~h ~\rm Mpc^{-1}$ at $z=6.5$, even in the absence of an excess radio background. This is also true for redshifts 6.8 and 7.2. On the other hand, the right panel shows that the upper limits at $z=7.8$ (and also for $z=8.2$ and 8.7) remain above the power spectra values from {\sc grizzly} for $A_r=0$.

As explained in Section~\ref{sec:like} the MCMC provides  the probability of the models to be excluded by the MWA upper limits. Each MCMC run consists of 10 walkers and $10^6$ steps. We find that the MCMC chains converge well before the final step.
The explored parameters ranges are listed in Table \ref{tab_source_param}. Note that all the priors are flat or uniform in log-scale. 

\begin{table}
\centering
\caption[]{Constraints on the excluded part of the IGM parameter space at different redshifts for the scenario where $A_r=0$, i.e. no excess radio background is considered in addition  to the CMB. These are obtained from the MCMC analysis using the multi-redshift MWA upper limits from \citet{2020MNRAS.493.4711T}. The priors are derived from the {\sc grizzly} simulations.}  
\small
\tabcolsep 4pt
\renewcommand\arraystretch{1.4}
   \begin{tabular}{| c | c c c c c |}
\hline

\makecell{IGM  \\ Parameters} & \makecell{Credible  \\ intervals}  & $z=6.5$ & $z=6.8$ & $z=7.2$ &   \\

\hline
\hline
\multirow{2}{4em}{$\overline{\XHII}$} 
& Prior & [0,  1] & [0, 1] & [0, 1]  & \\
\cline{2-6}

& 68$\%$ & [0, 0.15] & [0, 0.15] & [0, 0.26]  & \\
\cline{2-6}

& 95$\%$ & [0, 0.56] & [0, 0.53] & [0, 0.56] & \\

\hline
\hline
\multirow{2}{4em}{$\overline{\TK}$ (K)} 
& Prior & [1.17, $10^4$] & [1.28, $10^4$] & [1.38, $10^4$]  & \\ \cline{2-6}

& 68$\%$ & [1.17, 40.1] & [1.28, 11] & [1.38, 15.7]  & \\ \cline{2-6}

& 95$\%$ & [1.17, 251] & [1.28, 158] & [1.38, 90]  & \\

\hline
\hline
\multirow{2}{4em}{$f_{\rm heat}$} 
& Prior & [0, 1] & [0, 1] & [0, 1] & \\
\cline{2-6}

& 68$\%$ & [0, 0.28] & [0, 0.18] & [0, 0.28] & \\ 
\cline{2-6}

& 95$\%$ & [0, 0.63] & [0, 0.55] & [0, 0.58] & \\

\hline
\hline
\multirow{2}{4em}{$\overline{\TB}$ (mK)} 
& Prior & [-376, 23.3) & [-367, 23.8)  & [-360, 24.4)  & \\ 
\cline{2-6}

& 68$\%$ & [-376, -140] & [-367,-247] & [-360, -229]  & \\ 
\cline{2-6}

& 95$\%$ & [-376., -89] & [-367, -125] & [-360, -121] & \\

\hline
\hline
\multirow{2}{4em}{$R^{\rm heat}_{\rm peak}$ ($~h^{-1} \rm Mpc$)} 
& Prior & [0, 198] & [0, 203] & [0, 201]  & \\ 
\cline{2-6}

& 68$\%$ & [0, 4.7] & [0, 8.5] & [0, 18.2]  & \\ 
\cline{2-6}

& 95$\%$ & [0, 18.] & [0, 17.5] & [0, 30] & \\

\hline
\hline
\multirow{2}{4em}{$\Delta R^{\rm heat}_{\rm FWHM}$ ($~h^{-1} \rm Mpc$)} 
& Prior & [0, 440] & [0, 445] & [0, 452]  & \\ 
\cline{2-6}

& 68$\%$ & [0, 12.5] & [0, 16.6] & [0,35]  & \\ 
\cline{2-6}

& 95$\%$ & [0, 47] & [0,39] & [0, 63] & \\

\hline

\end{tabular}
\label{tab_mcmc_igmAr0}
\end{table}

To constrain the IGM parameters using our inference framework we consider a case with $A_r=0$, i.e. no excess radio background contribution to the CMB, and one in which we also vary the $A_r$ parameter. We will first explore the scenario without any excess radio background in Section \ref{sec:noRBG}, while  in Section \ref{sec:RBG} we present the constraints allowing $A_r$ parameter to vary.

\begin{figure*}
\begin{center}
\includegraphics[scale=0.46]{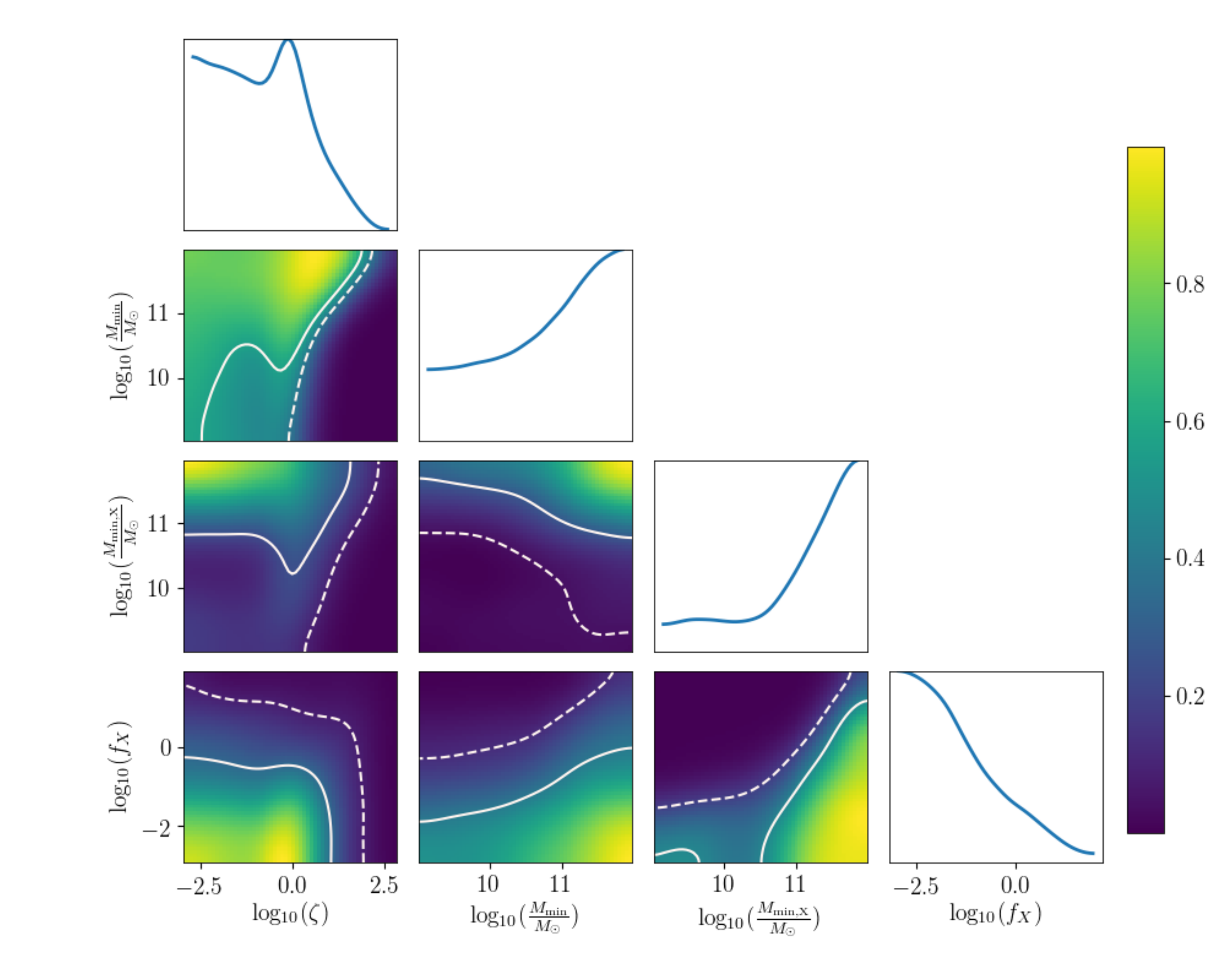}
    \caption{Posterior distribution of the simulation parameters at $z\approx 6.5$ from the MCMC analysis. Here $A_r=0$. The color-bar shows the marginalized probability of the models to be ruled out. The solid and dashed curves corresponds to the 68 and 95 per cent credible intervals of the models ruled out by the MWA upper limit at that redshift. The diagonal panels refer to the marginalized probability distributions for ruled out simulation parameters. }
   \label{image_sourceparam_z6.8_Ar0}
\end{center}
\end{figure*} 

\begin{figure*}
\begin{center}
\includegraphics[scale=0.46]{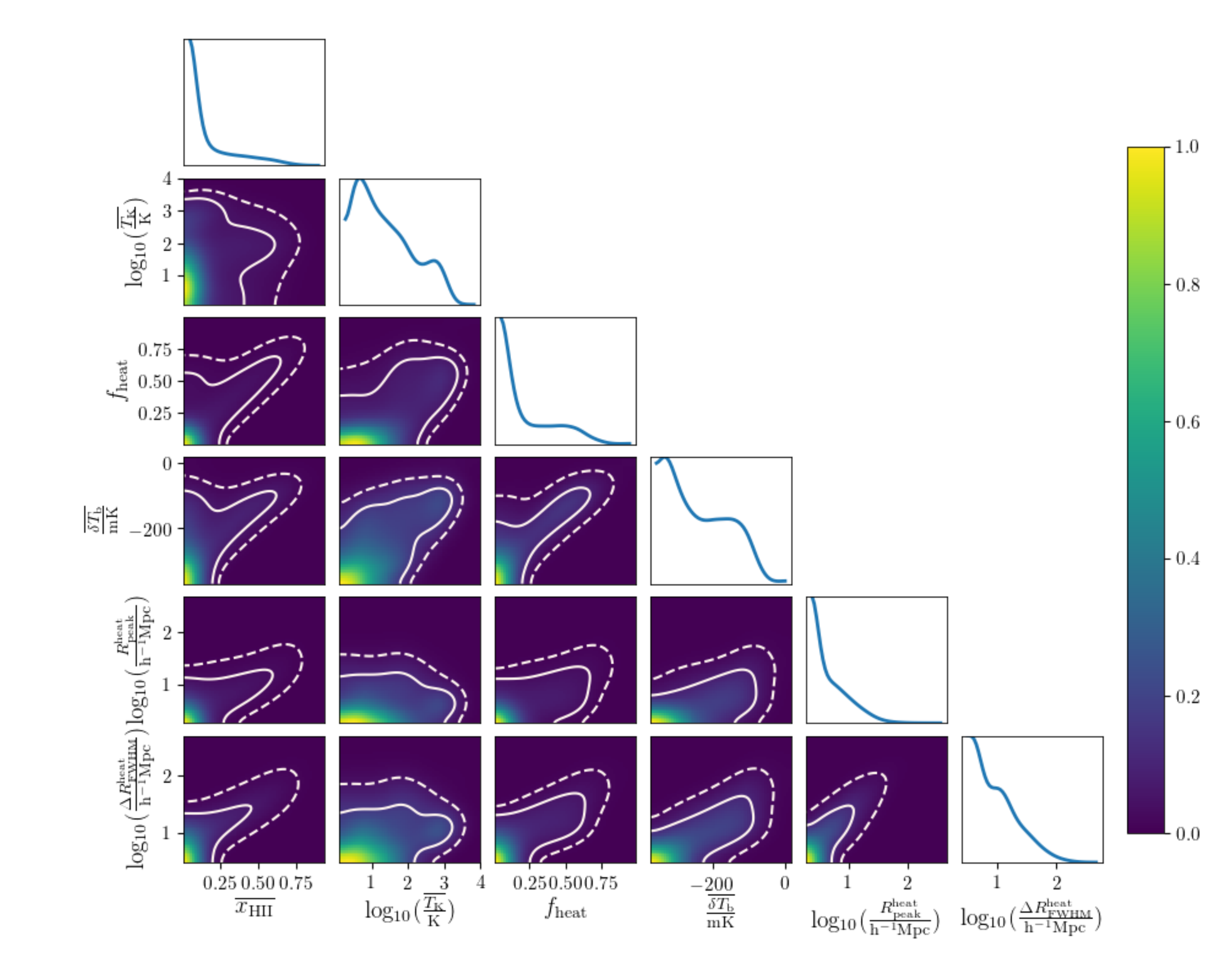}
    \caption{Posterior distribution of the IGM parameters at $z\approx 6.5$ from the MCMC analysis. 
    Here $A_r=0$. The color-bar shows the marginalized probability of the models to be ruled out. The solid and dashed curves corresponds to the 68 and 95 per cent credible intervals of the models ruled out by the MWA upper limit at that redshift. The diagonal panels refer to the marginalized probability distributions for ruled out IGM parameters.
    }
   \label{image_igmparam_z6.8_Ar0}
\end{center}
\end{figure*}

\subsection{Analysis without excess radio background}
\label{sec:noRBG}
First, we explore the four-dimensional source parameter space, i.e. $\zeta$, $M_{\rm min}$, $M_{\rm min, X}$ and $f_X$, using MCMC at six different redshifts. We fix $A_r=0$, i.e. we initially assume that the radio background is contributed by the CMB alone. 

The posterior distribution of the source parameter space of the excluded models at $z=6.5$, our fiducial redshift, is shown in Figure \ref{image_sourceparam_z6.8_Ar0}. Three main scenarios can be envisioned that produce a high amplitude of the large-scale power spectrum. In the first scenario the large-scale power spectrum is dominated by $\XHII$ fluctuations, which are enhanced by a distribution of rare but large \HII regions in the IGM.  Note that the large-scale power-spectrum decreases for a highly ionized IGM with $\overline{\XHII} \gtrsim 0.5$, and almost vanishes for $\overline{\XHII} \sim 1$.  This suggests that high values of $\zeta$ and low values of $M_{\rm min}$ should be disfavored among the excluded models, as confirmed in the figure. From this, we deduce that $\zeta \lesssim 2.4$ and $M_{\rm min} \gtrsim 3\times 10^{10} ~\MSUN$ are the most likely to be excluded within the 68 per cent credible intervals by the MWA upper limit at redshift 6.5 alone. Note also that a colder IGM, which is obtained with a smaller value of $f_X$, will also enhance the large-scale power spectrum in this scenario.  An alternative scenario that produces a high amplitude of the large-scale power spectrum is when the latter is dominated by $\TS$/$\TK$ fluctuations. This scenario can happen for a small number density of X-ray emitting sources, as numerous X-ray sources and high X-ray emissivity can easily heat the IGM above $\TCMB$ and suppress the large-scale power spectrum. Indeed, $f_X \lesssim 0.1$ and $M_{\rm min, X} \gtrsim 10^{11} ~\MSUN$ are the most likely to be excluded within the 68 per cent credible intervals by the MWA upper limit at redshift 6.5 alone. Alternatively, a scenario with neutral and cold IGM (as indicated by low values of $\zeta$ and $f_X$) where the large-scale power spectrum is dominated mainly by the density fluctuations is also important. 

While the parameters the $M_{\rm min}$, $M_{\rm min, X}$ and $f_X$ show a smooth distribution over the explored ranges, one can see a distinct peak around $\zeta \approx 1$ in the posterior distribution of $\zeta$ in Figure \ref{image_sourceparam_z6.8_Ar0}. The combination of $\zeta \approx 1$ and a high value of $M_{\rm min} \gtrsim 10^{11} ~\MSUN$ maximises the contribution from ionisation fluctuations. These ionization states correspond to $\overline{\XHII} \sim 0.5$. On the other hand, the other two scenarios, i.e. density fluctuations in a cold and neutral IGM and $\TS$ fluctuations in a neutral IGM, maximise the PDF for $\overline{\XHII}\sim 0$, which requires small values for $\zeta$. This bimodal feature in the PDF of $\zeta$ is consistent with the results of \citet{2020MNRAS.493.4728G}, where we perform a similar analysis for LOFAR upper limits at redshift 9.1. Note that we considered a prior on the ionization fraction at redshift 9.1 in \citet{2020MNRAS.493.4728G}.

\begin{table*}
\centering
\caption[]{Constraints on the excluded part of the IGM parameter space at different redshifts for a varying radio background scenario. These are obtained from the MCMC analysis using the multi-redshift upper limits from MWA \citep{2020MNRAS.493.4711T}. The priors are derived from the simulations run with {\sc grizzly}.} 
\small
\tabcolsep 8pt
\renewcommand\arraystretch{1.8}
   \begin{tabular}{| c | c  c c c c c c c |}
\hline

\makecell{IGM  \\ Parameters} & \makecell{Credible  \\ intervals}  & $z=6.5$ & $z=6.8$ & $z=7.2$ & $z=7.8$ & $z=8.2$ & $z=8.7$ &  \\

\hline
\hline
\multirow{2}{4em}{Log$_{10}$($A_r$)} 
& Prior & [-2, 2.6 ]  & [-2, 2.6 ] & [-2, 2.6 ]  & [-2, 2.6 ] & [-2, 2.6 ] & [-2, 2.6 ] & \\ 
\cline{2-9}

& 68$\%$ & [0.4, 2.6 ]  & [1, 2.6 ]  & [1.4, 2.6 ]  & [1.87, 2.6 ]  & [1.88, 2.6 ] & [1.9, 2.6 ] & \\ 
\cline{2-9}

& 95$\%$ & [-0.8, 2.6 ] & [-0.76, 2.6 ] & [-0.65, 2.6 ]  & [1.36, 2.6 ] & [1.38, 2.6 ]  & [1.43, 2.6 ] & \\

\hline
\hline
\multirow{2}{4em}{$\overline{\XHII}$} 
& Prior & [0,  1] & [0, 1] & [0, 1] & [0, 1] & [0, 1] & [0, 1] & \\
\cline{2-9}

& 68$\%$ & [0,  0.15] & [0, 0.15] & [0, 0.17] & [0, 0.12] & [0, 0.15] & [0, 0.12] & \\
\cline{2-9}

& 95$\%$ & [0, 0.58] & [0, 0.53] & [0, 0.56] & [0, 0.56] & [0, 0.55] & [0, 0.52] & \\

\hline
\hline
\multirow{2}{4em}{$\overline{\TK}$ (K)} 
& Prior & [1.17, $10^4$] & [1.28, $10^4$] & [1.38, $10^4$] & [1.63, $10^4$] & [1.75, $10^4$] & [2, $10^4$] & \\ \cline{2-9}

& 68$\%$ & [1.17, 208] & [1.28, 80.7] & [1.38, 69] & [1.63, 69] & [1.75, 68] & [2, 42] & \\ \cline{2-9}

& 95$\%$ & [1.17, 1025] & [1.28, 795] & [1.38, 790] & [1.63, 602] & [1.75, 592] & [2, 400.] & \\

\hline
\hline
\multirow{2}{4em}{$f_{\rm heat}$} 
& Prior & [0,  1] & [0, 1] & [0, 1] & [0, 1] & [0, 1] & [0, 1] & \\
\cline{2-9}

& 68$\%$ & [0, 0.33] & [0, 0.28] & [0, 0.28] & [0, 0.28] & [0, 0.23] & [0, 0.18] & \\ 
\cline{2-9}

& 95$\%$ & [0, 0.64] & [0, 0.6] & [0, 0.59] & [0, 0.56] & [0, 0.57] & [0, 0.55] & \\

\hline
\hline

\multirow{2}{4em}{Log$_{10}$($\frac{|\overline{\TB}|}{\rm mK}$)} 
& Prior & [1.37, 4.6] & [1.38, 4.59]  & [1.39, 4.58]  & [1.4, 4.57]  & [1.41, 4.56]  & [1.43, 4.54]  & \\ 
\cline{2-9}

& 68$\%$ & [2.18, 3.3] & [2.4, 3.57] & [2.65, 3.76] & [3.21, 4] & [3.24, 4.04] & [3.38, 4.06] & \\ 
\cline{2-9}

& 95$\%$ & [1.96, 4.2] & [2.18, 4.24] & [2.29, 4.27] & [2.94, 4.39] & [2.98, 4.38] & [3.19, 4.43] & \\

\hline
\hline
\multirow{2}{4em}{$R^{\rm heat}_{\rm peak}$ ($~h^{-1} \rm Mpc$)} 
& Prior & [0, 198] & [0, 203] & [0, 201] & [0, 201] & [0, 216] & [0, 198] & \\ 
\cline{2-9}

& 68$\%$ & [0, 4.] & [0, 3.3] & [0, 3] & [0,3.6] & [0,3.6] & [0,3.5] & \\ 
\cline{2-9}

& 95$\%$ & [0, 12.] & [0, 10] & [0, 13] & [0,12] & [0,11.5] & [0, 11.5] & \\

\hline
\hline
\multirow{2}{4em}{$\Delta R^{\rm heat}_{\rm FWHM}$ ($~h^{-1} \rm Mpc$)} 
& Prior & [0, 440] & [0, 445] & [0, 452] & [0, 452] & [0, 447] & [0, 452] & \\ 
\cline{2-9}

& 68$\%$ & [0, 14] & [0, 9.1] & [0, 11.4] & [0, 10.] & [0, 11.2] & [0, 9.3] & \\ 
\cline{2-9}

& 95$\%$ & [0, 29] & [0, 22] & [0, 26] & [0, 23] & [0, 25] & [0, 23] & \\
\hline
\end{tabular}
\label{tab_mcmc_igmAr}
\end{table*}

Figure \ref{image_igmparam_z6.8_Ar0} shows the posterior distributions of the IGM parameters at redshift 6.5. The constraints on such parameters are also listed in Table \ref{tab_mcmc_igmAr0}. We find that an IGM with $\overline{\XHII} \lesssim 0.56$,  $\overline{\TK} \lesssim 250$ K, $f_{\rm heat}\lesssim 0.63$, $ R^{\rm heat}_{\rm peak}\lesssim 18 ~h^{-1}$ Mpc and $ \Delta R^{\rm heat}_{\rm FWHM}\lesssim 50~h^{-1}$ Mpc falls within 95 per cent credible intervals of the ruled out IGM scenarios. The same limit for the average brightness temperature is $ \overline{\TB} \lesssim -90 $ mK. The PDFs of $\overline{\XHII}$, $f_{\rm heat}$ and $ \overline{\TB}$ suggest that the set of excluded models contains a significant number of models in which the IGM is still neutral and cold at redshift $z\approx 6.5$. A similar conclusion was reached by \citet[]{2020arXiv200802639G}. However, it is hard to tell from the achieved limits on the IGM parameters if these neutral models are completely cold or contain patchy heated regions in an otherwise cold IGM.

The 68 and 95 per cent credible intervals limits of the excluded IGM parameters for other redshifts are also listed in  Table \ref{tab_mcmc_igmAr0}. As the upper limit of the power spectrum from MWA is weaker at a higher redshift, the limits also become weaker, and especially for $z\gtrsim 7.8$ their significance is very small. 
In fact, we neither get any constraint on the source nor on the IGM parameters at $z\approx 8.7$.
For $z\approx 7.8$ and 8.2, the values of log$_{10}$($\mathcal{L}_{\rm ex}$) remain $\lesssim -5$, which corresponds to the value of log$_{10}$($\mathcal{L}_{\rm ex}$) when a modelled power spectrum is below 3$\sigma$ error of the observed upper limits at all three scales. 
Clearly, the reason is that all the {\sc grizzly} models have power spectra values that are well below the upper limits (see Figure \ref{image_psts_z}). As the significance of the limits at redshifts $\gtrsim 7.8$ is poor, we do not show them in this work. Note that the limits as stated in Table \ref{tab_mcmc_igmAr0} only stand for the part of the parameter space which has the largest probability to be ruled out.


\subsection{Analysis with varying radio background}
\label{sec:RBG}
Now, we explore the five-dimensional simulation parameter space, i.e. $\zeta$, $M_{\rm min}$, $M_{\rm min, X}$, $f_X$ and $A_r$ using MCMC at six different redshifts. The explored ranges of the simulation parameters are listed in Table \ref{tab_source_param}.

\begin{figure*}
\begin{center}
\includegraphics[scale=0.46]{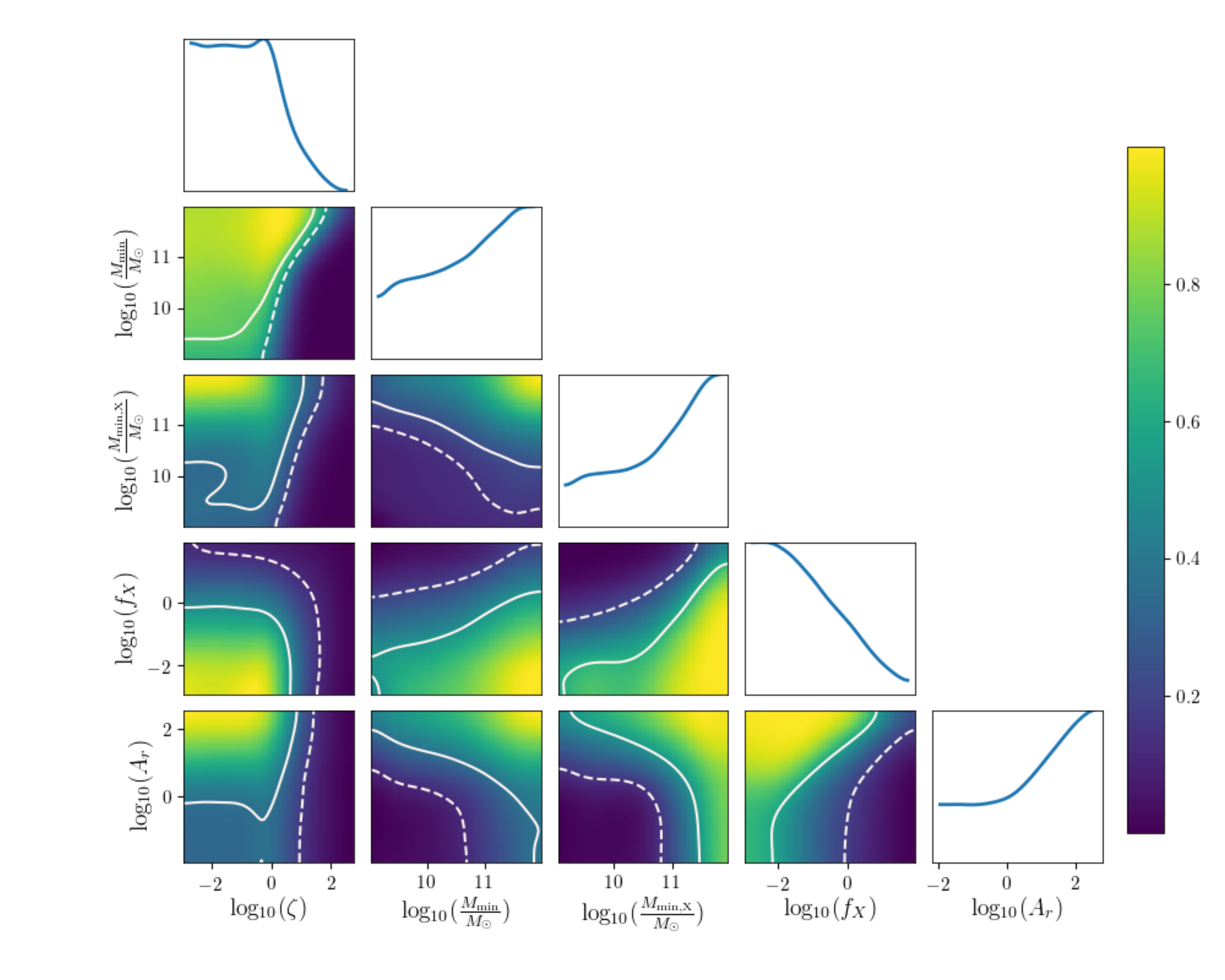}
    \caption{Posterior distribution of the simulation parameters at $z\approx 6.5$ from the MCMC analysis. The color-bar shows the marginalized probability of the models to be ruled out. The solid and dashed curves corresponds to the 68 and 95 per cent credible intervals of the models ruled out by the MWA upper limit at that redshift. The diagonal panels refer to the marginalized probability distributions for ruled out simulation parameters.  }
   \label{image_sourceparam_z6.8}
\end{center}
\end{figure*}

Figure \ref{image_sourceparam_z6.8} shows the posterior distribution of the simulation parameters of the models with high exclusion probability at $z \approx 6.5$. The 68 per cent credible intervals limits of these model parameters are  $\zeta \lesssim 1$, $M_{\rm min} \gtrsim 2\times 10^{10} ~\MSUN$, $M_{\rm min, X} \gtrsim 4\times 10^{10} ~\MSUN$ and $f_X \lesssim 0.1$. As the amplitude of the power spectrum increases with increasing $A_r$, the excluded models prefer a higher value of $A_r$. From this five-dimensional MCMC analysis, in fact, we find that $A_r \gtrsim 0.15$ ($ \gtrsim 2.5$) is the most likely to be excluded within the 95 (68) per cent credible intervals by the MWA upper limit at redshift 6.5 alone.

\begin{figure*}
\begin{center}
\includegraphics[scale=0.46]{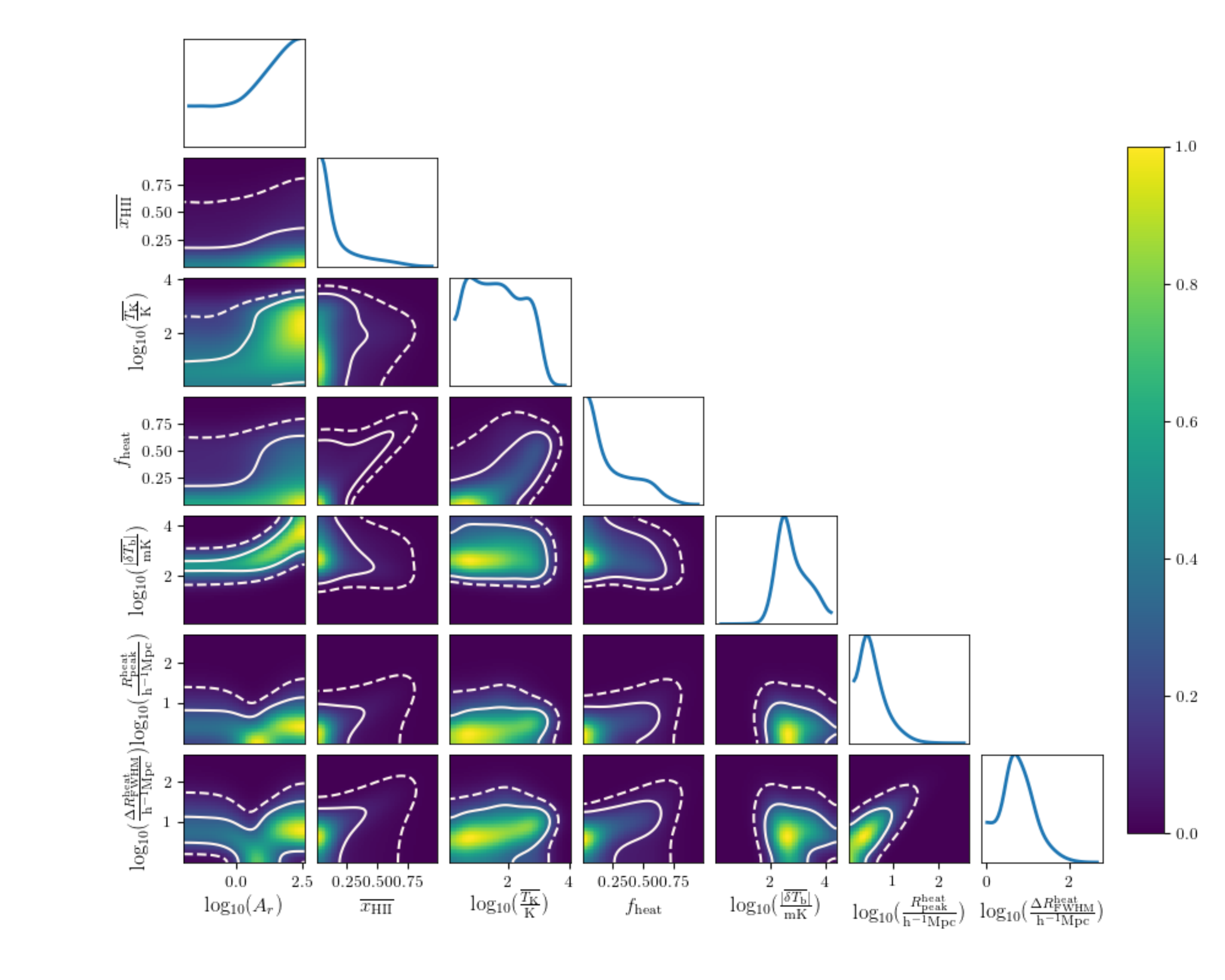}
    \caption{Posterior distribution of the IGM parameters at $z\approx 6.5$ from the MCMC analysis. The color-bar shows the marginalized probability of the models to be ruled out. The solid and dashed curves corresponds to the 68 and 95 per cent credible intervals of the models ruled out by the MWA upper limit at that redshift. The diagonal panels refer to the marginalized probability distributions for ruled out IGM parameters. Note that we have used Logarithm of the absolute values of $\overline{\TB}$ as the range is large. In reality, $\overline{\TB}$ values of these excluded models are mostly negative. }
   \label{image_igmparam_z6.8}
\end{center}
\end{figure*}

\begin{figure*}
\begin{center}
\includegraphics[scale=0.37]{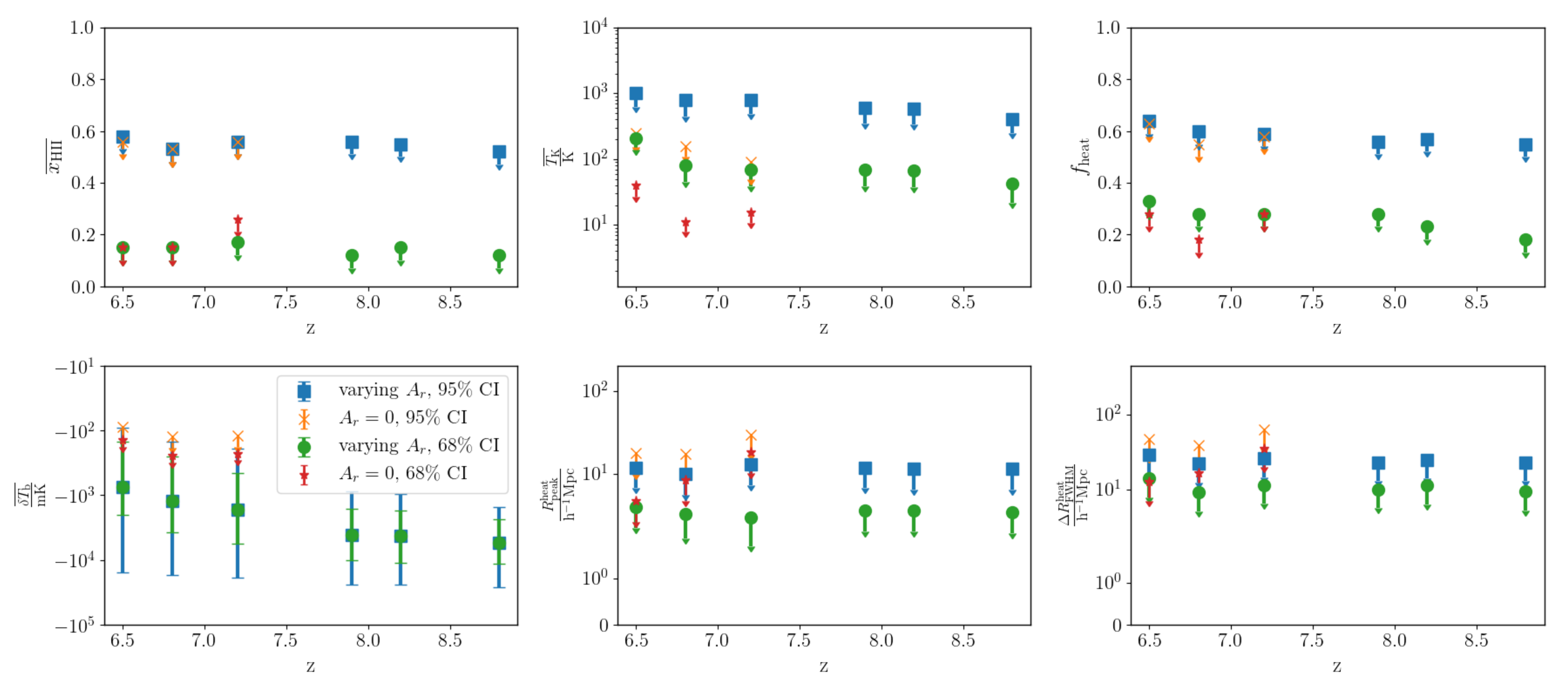}
    \caption{limits on the IGM parameters as obtained from the excluded models at six different redshifts using the recent MWA upper limits within 68 and 95 per cent credible interval. These constraints have been determined at each redshift independently. For $z\ge 7.8$, no constraints are found in absence of an additional radio background ($A_r=0)$.}
   \label{image_MWA_con}
\end{center}
\end{figure*}

Figure \ref{image_igmparam_z6.8} shows instead the posterior distributions of the IGM parameters at the same redshift. These constraints are also listed in Table \ref{tab_mcmc_igmAr} and show that an IGM with $\overline{\XHII} \lesssim 0.6$,  $\overline{\TK} \lesssim 10^3$ K, $f_{\rm heat}\lesssim 0.65$, $ R^{\rm heat}_{\rm peak}\lesssim 13 ~h^{-1}$ Mpc and $ \Delta R^{\rm heat}_{\rm FWHM}\lesssim 30~h^{-1}$ Mpc falls within 95 per cent credible intervals of the ruled out IGM scenarios. As these intervals indicate a larger value of $A_r$ (i.e. $A_r\gtrsim 2.5$), the derived limit of the average brightness temperature, $\overline{\TB} \gtrsim -10^4$ mK, is weaker compared to what we had found for $A_r=0$.

The constraints on the IGM parameters for the other five redshifts are also listed in Table \ref{tab_mcmc_igmAr} and presented in Fig. \ref{image_MWA_con}. We find that the 95 per cent credible intervals for each of the quantities, $\overline{\XHII}$, $f_{\rm heat}$, $ R^{\rm heat}_{\rm peak}$ and $ \Delta R^{\rm heat}_{\rm FWHM}$, are similar at all six redshifts.  This shows that the ionization and thermal states of the IGM which are given by the IGM parameter values within these limits produce extreme amplitudes of the large-scale power spectra, while the limits on the IGM parameters are independent of redshift. Note that because the source populations differ between redshifts, it will require different values of the source parameters to achieve similar thermal and ionization state of the IGM at different redshifts. As the power spectrum amplitude increases with increasing $A_r$ and the upper limits are weaker at higher $z$, the 95 per cent credible interval limit for $A_r$ increases as well, for example from $\approx 0.15$ at redshift 6.5 to $\approx27$ at redshift 8.7. These correspond to $\TEFF \approx 26$ K and 140 K at redshifts 6.5 and 8.7, respectively.  The limit at redshift 6.5 is equivalent to an excess radio background which is $0.008\%$  of the CMB at 1.42 GHz. Excess of $0.009\%, 0.01\%, 1.2\%, 1.3\%$ and $1.5\%$ are found from the limits at redshifts 6.8, 7.2, 7.8, 8.2 and 8.7, respectively.


\section{Discussion}
\label{sec:discussion}
We highlight that the constraints on the IGM and the simulation parameters obtained in the previous section do not correspond to models ruled out with a high significance, but rather to models that have a high probability to be excluded. Future stronger upper limits will improve the significance/probability of the models to be ruled out.

It should be noted that the part of the parameter space which has high probability to be excluded is associated with models that produce a high amplitude of the large-scale power spectrum.  Large ionized regions in a cold IGM, and/or large heated regions in a non-uniform spin temperature IGM are the obvious scenarios that come to mind. Alternatively, density fluctuations in a cold IGM can also produce high values of the large-scale power spectrum, especially at the lower redshifts where the density fluctuations are larger and the IGM is expected to be colder in the absence of heating sources. The same models also suggest a high amplitude of the excess radio background, as this boosts the fluctuations at all scales. It is challenging to devise scenarios alternative to those discussed here that have such a high chance of exclusion without accounting for other non-standard physics, such as e.g. an excess cooling mechanism that makes the gas in the IGM colder than in a standard scenario. 

As we consider each redshift independently, the models more likely to be excluded show similar IGM conditions (e.g. $\overline{\XHII} \lesssim 0.5$ and cold IGM) at all redshifts, although the constraints become obviously stronger at redshifts with tighter upper limits. In our previous study \citep{2020MNRAS.493.4728G}, we obtained constraints on the IGM parameters at $z \approx 9.1$ using the upper limits from LOFAR \citep{2020MNRAS.493.1662M}, finding that an IGM with $f_{\rm heat} \lesssim 0.35$, $\TK \lesssim 160$ K, $R^{\rm heat}_{\rm peak} \lesssim 70 ~h^{-1}\rm Mpc$ and $\Delta R^{\rm heat}_{\rm FWHM} \lesssim 110 ~h^{-1} \rm Mpc$ is the most likely to be ruled out within 95 per cent credible interval limits. This physical state of the IGM is similar to those found to be more likely ruled out in the present study. This is to be expected, as those states of the IGM which produce high amplitudes of the large-scale power spectra are easy to exclude by the 21-cm signal upper limits. 

We note that, for some of the models with high exclusion probability, the large-scale power spectrum is driven by $\TS$ fluctuations (and/or density fluctuations) only when the $\XHII$ fluctuations are small. These models have an ionization fraction  $\overline{\XHII} \approx 0$ at $z \approx 6.5$, which is inconsistent with the combined constraints from observations of the Thomson scattering optical depth and $z>6$ quasar spectra \citep[see e.g.,][]{Mitra15}, as well as $\lya$ emitter statistics \citep{2010ApJ...723..869O}, spectroscopy of gamma-ray bursts  \citep{2013ApJ...774...26C}, dark pixel statistics of the spectroscopy of high-$z$ quasars \citep{2015MNRAS.447..499M}, which all indicate that the IGM is not highly neutral at redshift $\lesssim 7$.

The constraints presented in the previous section are obtained analysing only one of the six upper limits/redshifts at a time. In the MCMC analysis the upper limits are not all combined into a single likelihood, as otherwise this would be mostly driven by the best upper limit, i.e. the one at $z\approx 6.5$, biasing  the conclusions on the IGM properties at individual redshifts. 

A recent study by \citet[]{2020arXiv200802639G} obtained constraints on source parameters using the MWA upper limits by combining all redshifts into a single likelihood for a set of redshift independent source parameters. The derived constraints on the IGM parameters at different redshifts as obtained in fig 6 of \citet[]{2020arXiv200802639G} depend on the background source model. As a result, the constraints on $\overline{\XHI}$ follow a redshift evolution as expected in such model. We on the other hand constrain the IGM state at each redshift using the upper limits of the power spectra at that redshift only, and thus the constraints  do not depend on the redshift evolution in the background model. The differences in the analysis methods make it hard to directly compare our results with those in \citet[]{2020arXiv200802639G}. However, as the results of \citet[]{2020arXiv200802639G} are expected to be biased by the strongest upper limits at redshift 6.5, one might expect that the IGM states of the excluded models of both studies are qualitatively similar at redshift 6.5, which we indeed have found to be the case in Section~\ref{sec:noRBG}.

Using LOFAR upper limits at $z=9.1$ from \citet{2020MNRAS.493.1662M}, \citet{2020MNRAS.498.4178M} obtained constraints on the excess radio background as $0.1-9.6\%$ of the CMB at 1.42 GHz. Note that these 95 per cent credible limits are obtained from marginalizing models {\it which are valid} under the LOFAR upper limits. Our limits are instead derived from the marginalization of the {\it excluded models}. It is therefore not possible to compare the above constraint to the one we find, namely that the 95 per cent credible limits of $A_r$ at redshift 8.7 (6.5) correspond to an excess radio background which is $1.5\%$ ($0.008\%$) of the CMB at 1.42 GHz.

\section{Summary \& Conclusions}
\label{sec:con}

\citet{2020MNRAS.493.4711T} has provided new upper limits on the dimensionless spherically averaged power spectrum of the 21-cm signal from MWA at six different redshifts, in the range  $6.5-8.7$. The best 2$\sigma$ upper limit from that paper, obtained from 110 hours of integration time, is $\Delta^2(k=0.14~h ~\rm Mpc^{-1})=(43)^2$ mK$^2$ at $z\approx 6.5$. Here we have used these limits to rule out possible reionization scenarios. The main focus of this study is to constrain the ionization and thermal states of the IGM.

We use the {\sc grizzly} code to generate hundred of thousands of models for different combinations of parameters, namely, ionization efficiency ($\zeta$), minimum mass of the UV emitting halos ($M_{\rm min}$), minimum mass of X-ray emitting halos ($M_{\rm min, X}$), X-ray heating efficiency ($f_X$) and excess radio background efficiency ($A_r$). The outputs of these models are power spectra and derived IGM parameters,
such as volume averaged ionization fraction ($\overline{\XHII}$), volume averaged gas temperature of the partially ionized IGM ($\overline{\TK}$), mass averaged brightness temperature ($\overline{\TB}$), volume fraction of the heated region ($f_{\rm heat}$), size of the heated regions at which the PDF of the sizes peaks ($R^{\rm heat}_{\rm peak})$ and the FWHM of the PDFs ($\Delta R^{\rm heat}_{\rm FWHM})$. The outputs from the {\sc grizzly} simulations are used within an MCMC framework to constrain the simulation parameters using the upper limits from MWA. The simulation parameters from the MCMC chains are later used to generate lists of IGM parameters. 
The main findings of the paper are the followings.

\begin{itemize}
\item From the set of power spectra generated using {\sc grizzly}, we do not find any model which can be excluded in the absence of an extra radio background component at redshifts 7.8, 8.2 and 8.7. On the other hand, we find a significant number of models that can be ruled out at the other three redshifts. 

\item From the MCMC analysis, we find that the excluded models are more likely to have a smaller number density of the UV as well as X-ray emitting sources. At redshift $\approx 6.5$, for 68 per cent credible region of the excluded models, the mass of the dark matter halo that hosts these sources is $ \gtrsim 3 \times 10^{10} ~\MSUN$, while $\zeta \lesssim 1.4$ and $f_X \lesssim 0.1$.

\item The probability with which a reionization model can be excluded increases with increasing $A_r$. We found $A_r \gtrsim 0.15$  for the 95 per cent credibility interval of the excluded models at redshift 6.8. This limit increases for higher redshifts as the upper limits are weaker there. At the highest redshift of $z=8.7$  we find $A_r \gtrsim 27$ for the same credibility interval. Within 95 per cent credible interval limits, the additional radio background is at least $0.008\%, 0.009\%, 0.01\%, 1.2\%, 1.3\%$ and $1.5\%$ of the CMB at 1.42 GHz at redshifts 6.5, 6.8, 7.2, 7.8, 8.2 and 8.7 respectively. These values differ with redshifts because the upper limits have been considered individually. In the presence of such radio background, the limits on the IGM parameters for the 95 per cent credible interval of the excluded models are  $\overline{\XHII} \lesssim 0.6$, $\overline{\TK} \lesssim 10^3$ K, $f_{\rm heat} \lesssim 0.6$, $R^{\rm heat}_{\rm peak} \lesssim 10 ~ h^{-1}~ \rm Mpc$, $\Delta R^{\rm heat}_{\rm FWHM} \lesssim 30 ~h^{-1} ~\rm Mpc$.

\end{itemize}

The limits on the IGM parameters are similar at all redshifts considered in this study, indicating that certain IGM states correspond to large amplitudes of the large-scale power spectra and thus are more easily ruled out by the upper limits on the 21-cm signal power spectrum. More stringent upper limits will provide stronger constraints on the IGM parameters. In addition, combining these with other observational constraints such as the observation of the global 21-cm signal and high-$z$ galaxies will strengthen the understanding of the IGM during the EoR.

\section*{Acknowledgements}
We would like to thank the anonymous referee for insightful comments. We thank Cathryn M. Trott for providing the lists of the MWA upper limits and the corresponding errors, and Bradley Greig for useful discussion related to this work. 
We acknowledge that the results in this paper have been achieved using the PRACE Research Infrastructure resources Curie based at the Tr$\grave{\rm e}$s Grand Centre de Calcul (TGCC) operated by CEA near Paris, France and Marenostrum based in the Barcelona Supercomputing Center, Spain. Time on these resources was awarded by PRACE under PRACE4LOFAR grants 2012061089 and 2014102339 as well as under the Multi-scale Reionization grants 2014102281 and 2015122822. The computations were partly enabled by resources provided by the Swedish National Infrastructure for Computing (SNIC) at PDC partially funded by the Swedish Research Council through grant agreement no. 2018-05973. RG and SZ furthermore acknowledge support by the Israel Science Foundation (grant no. 255/18). GM is supported by Swedish Research Council grants 2016-03581 and 2020-04691.

\section*{DATA AVAILABILITY}

The derived data generated in this research will be shared on reasonable request to the corresponding authors.

\bibliography{mybib}

\begin{thebibliography}{}
\makeatletter
\relax
\def\mn@urlcharsother{\let\do\@makeother \do\$\do\&\do\#\do\^\do\_\do\%\do\~}
\def\mn@doi{\begingroup\mn@urlcharsother \@ifnextchar [ {\mn@doi@}
  {\mn@doi@[]}}
\def\mn@doi@[#1]#2{\def\@tempa{#1}\ifx\@tempa\@empty \href
  {http://dx.doi.org/#2} {doi:#2}\else \href {http://dx.doi.org/#2} {#1}\fi
  \endgroup}
\def\mn@eprint#1#2{\mn@eprint@#1:#2::\@nil}
\def\mn@eprint@arXiv#1{\href {http://arxiv.org/abs/#1} {{\tt arXiv:#1}}}
\def\mn@eprint@dblp#1{\href {http://dblp.uni-trier.de/rec/bibtex/#1.xml}
  {dblp:#1}}
\def\mn@eprint@#1:#2:#3:#4\@nil{\def\@tempa {#1}\def\@tempb {#2}\def\@tempc
  {#3}\ifx \@tempc \@empty \let \@tempc \@tempb \let \@tempb \@tempa \fi \ifx
  \@tempb \@empty \def\@tempb {arXiv}\fi \@ifundefined
  {mn@eprint@\@tempb}{\@tempb:\@tempc}{\expandafter \expandafter \csname
  mn@eprint@\@tempb\endcsname \expandafter{\@tempc}}}

\bibitem[\protect\citeauthoryear{{Ba{\~n}ados} et~al.,}{{Ba{\~n}ados}
  et~al.}{2018}]{2018Natur.553..473B}
{Ba{\~n}ados} E.,  et~al., 2018, \mn@doi [\nat] {10.1038/nature25180}, \href
  {https://ui.adsabs.harvard.edu/\#abs/2018Natur.553..473B} {553, 473}

\bibitem[\protect\citeauthoryear{{Barkana}}{{Barkana}}{2018}]{2018Natur.555...71B}
{Barkana} R.,  2018, \mn@doi [\nat] {10.1038/nature25791}, \href
  {http://adsabs.harvard.edu/abs/2018Natur.555...71B} {555, 71}

\bibitem[\protect\citeauthoryear{{Barkana} \& {Loeb}}{{Barkana} \&
  {Loeb}}{2005}]{barkana05b}
{Barkana} R.,  {Loeb} A.,  2005, \mn@doi [\apj] {10.1086/429954}, \href
  {http://adsabs.harvard.edu/abs/2005ApJ...626....1B} {626, 1}

\bibitem[\protect\citeauthoryear{{Barry}, {Hazelton}, {Sullivan}, {Morales}  \&
  {Pober}}{{Barry} et~al.}{2016}]{2016MNRAS.461.3135B}
{Barry} N.,  {Hazelton} B.,  {Sullivan} I.,  {Morales} M.~F.,   {Pober} J.~C.,
  2016, \mn@doi [\mnras] {10.1093/mnras/stw1380}, \href
  {https://ui.adsabs.harvard.edu/abs/2016MNRAS.461.3135B} {461, 3135}

\bibitem[\protect\citeauthoryear{{Barry} et~al.,}{{Barry}
  et~al.}{2019}]{2019ApJ...884....1B}
{Barry} N.,  et~al., 2019, \mn@doi [\apj] {10.3847/1538-4357/ab40a8}, \href
  {https://ui.adsabs.harvard.edu/abs/2019ApJ...884....1B} {884, 1}

\bibitem[\protect\citeauthoryear{{Behroozi} \& {Silk}}{{Behroozi} \&
  {Silk}}{2015}]{2015ApJ...799...32B}
{Behroozi} P.~S.,  {Silk} J.,  2015, \mn@doi [\apj]
  {10.1088/0004-637X/799/1/32}, \href
  {https://ui.adsabs.harvard.edu/abs/2015ApJ...799...32B} {799, 32}

\bibitem[\protect\citeauthoryear{{Bonaldi} \& {Brown}}{{Bonaldi} \&
  {Brown}}{2015}]{2015MNRAS.447.1973B}
{Bonaldi} A.,  {Brown} M.~L.,  2015, \mn@doi [\mnras] {10.1093/mnras/stu2601},
  \href {https://ui.adsabs.harvard.edu/abs/2015MNRAS.447.1973B} {447, 1973}

\bibitem[\protect\citeauthoryear{{Bowman} \& {Rogers}}{{Bowman} \&
  {Rogers}}{2010}]{2010Natur.468..796B}
{Bowman} J.~D.,  {Rogers} A.~E.~E.,  2010, \mn@doi [\nat]
  {10.1038/nature09601}, \href
  {http://adsabs.harvard.edu/abs/2010Natur.468..796B} {468, 796}

\bibitem[\protect\citeauthoryear{{Bowman}, {Rogers}, {Monsalve}, {Mozdzen}  \&
  {Mahesh}}{{Bowman} et~al.}{2018}]{EDGES2018}
{Bowman} J.~D.,  {Rogers} A.~E.~E.,  {Monsalve} R.~A.,  {Mozdzen} T.~J.,
  {Mahesh} N.,  2018, \mn@doi [\nat] {10.1038/nature25792}, \href
  {http://adsabs.harvard.edu/abs/2018Natur.555...67B} {555, 67}

\bibitem[\protect\citeauthoryear{{Bradley}, {Tauscher}, {Rapetti}  \&
  {Burns}}{{Bradley} et~al.}{2019}]{2019ApJ...874..153B}
{Bradley} R.~F.,  {Tauscher} K.,  {Rapetti} D.,   {Burns} J.~O.,  2019, \mn@doi
  [\apj] {10.3847/1538-4357/ab0d8b}, \href
  {https://ui.adsabs.harvard.edu/abs/2019ApJ...874..153B} {874, 153}

\bibitem[\protect\citeauthoryear{{Chapman}, {Zaroubi}, {Abdalla}, {Dulwich},
  {Jeli{\'c}}  \& {Mort}}{{Chapman} et~al.}{2016}]{2016MNRAS.458.2928C}
{Chapman} E.,  {Zaroubi} S.,  {Abdalla} F.~B.,  {Dulwich} F.,  {Jeli{\'c}} V.,
   {Mort} B.,  2016, \mn@doi [\mnras] {10.1093/mnras/stw161}, \href
  {https://ui.adsabs.harvard.edu/abs/2016MNRAS.458.2928C} {458, 2928}

\bibitem[\protect\citeauthoryear{{Cheng} et~al.,}{{Cheng}
  et~al.}{2018}]{2018ApJ...868...26C}
{Cheng} C.,  et~al., 2018, \mn@doi [\apj] {10.3847/1538-4357/aae833}, \href
  {https://ui.adsabs.harvard.edu/abs/2018ApJ...868...26C} {868, 26}

\bibitem[\protect\citeauthoryear{{Chornock}, {Berger}, {Fox}, {Lunnan},
  {Drout}, {Fong}, {Laskar}  \& {Roth}}{{Chornock}
  et~al.}{2013}]{2013ApJ...774...26C}
{Chornock} R.,  {Berger} E.,  {Fox} D.~B.,  {Lunnan} R.,  {Drout} M.~R.,
  {Fong} W.-f.,  {Laskar} T.,   {Roth} K.~C.,  2013, \mn@doi [\apj]
  {10.1088/0004-637X/774/1/26}, \href
  {https://ui.adsabs.harvard.edu/abs/2013ApJ...774...26C} {774, 26}

\bibitem[\protect\citeauthoryear{{Cohen}, {Fialkov}, {Barkana}  \&
  {Monsalve}}{{Cohen} et~al.}{2020}]{2020MNRAS.495.4845C}
{Cohen} A.,  {Fialkov} A.,  {Barkana} R.,   {Monsalve} R.~A.,  2020, \mn@doi
  [\mnras] {10.1093/mnras/staa1530}, \href
  {https://ui.adsabs.harvard.edu/abs/2020MNRAS.495.4845C} {495, 4845}

\bibitem[\protect\citeauthoryear{{Datta}, {Bharadwaj}  \& {Choudhury}}{{Datta}
  et~al.}{2007}]{kanan2007MNRAS.382..809D}
{Datta} K.~K.,  {Bharadwaj} S.,   {Choudhury} T.~R.,  2007, \mn@doi [\mnras]
  {10.1111/j.1365-2966.2007.12421.x}, \href
  {http://adsabs.harvard.edu/abs/2007MNRAS.382..809D} {382, 809}

\bibitem[\protect\citeauthoryear{{Datta}, {Bowman}  \& {Carilli}}{{Datta}
  et~al.}{2010}]{2010ApJ...724..526D}
{Datta} A.,  {Bowman} J.~D.,   {Carilli} C.~L.,  2010, \mn@doi [\apj]
  {10.1088/0004-637X/724/1/526}, \href
  {http://adsabs.harvard.edu/abs/2010ApJ...724..526D} {724, 526}

\bibitem[\protect\citeauthoryear{{Davies} et~al.,}{{Davies}
  et~al.}{2018}]{2018ApJ...864..142D}
{Davies} F.~B.,  et~al., 2018, \mn@doi [\apj] {10.3847/1538-4357/aad6dc}, \href
  {https://ui.adsabs.harvard.edu/abs/2018ApJ...864..142D} {864, 142}

\bibitem[\protect\citeauthoryear{{Dawoodbhoy} et~al.,}{{Dawoodbhoy}
  et~al.}{2018}]{2018MNRAS.480.1740D}
{Dawoodbhoy} T.,  et~al., 2018, \mn@doi [\mnras] {10.1093/mnras/sty1945}, \href
  {https://ui.adsabs.harvard.edu/abs/2018MNRAS.480.1740D} {480, 1740}

\bibitem[\protect\citeauthoryear{{DeBoer} et~al.,}{{DeBoer}
  et~al.}{2017}]{2017PASP..129d5001D}
{DeBoer} D.~R.,  et~al., 2017, \mn@doi [Publications of the Astronomical
  Society of the Pacific] {10.1088/1538-3873/129/974/045001}, \href
  {https://ui.adsabs.harvard.edu/\#abs/2017PASP..129d5001D} {129, 045001}

\bibitem[\protect\citeauthoryear{Dixon, Iliev, Mellema, Ahn  \& Shapiro}{Dixon
  et~al.}{2016}]{Dixon2016TheReionization}
Dixon K.~L.,  Iliev I.~T.,  Mellema G.,  Ahn K.,   Shapiro P.~R.,  2016,
  \mn@doi [Monthly Notices of the Royal Astronomical Society]
  {10.1093/mnras/stv2887}, 456, 3011

\bibitem[\protect\citeauthoryear{{Dowell} \& {Taylor}}{{Dowell} \&
  {Taylor}}{2018}]{2018ApJ...858L...9D}
{Dowell} J.,  {Taylor} G.~B.,  2018, \mn@doi [\apjl]
  {10.3847/2041-8213/aabf86}, \href
  {https://ui.adsabs.harvard.edu/abs/2018ApJ...858L...9D} {858, L9}

\bibitem[\protect\citeauthoryear{{Draine} \& {Miralda-Escud{\'e}}}{{Draine} \&
  {Miralda-Escud{\'e}}}{2018}]{2018ApJ...858L..10D}
{Draine} B.~T.,  {Miralda-Escud{\'e}} J.,  2018, \mn@doi [\apjl]
  {10.3847/2041-8213/aac08a}, \href
  {https://ui.adsabs.harvard.edu/abs/2018ApJ...858L..10D} {858, L10}

\bibitem[\protect\citeauthoryear{{Eastwood} et~al.,}{{Eastwood}
  et~al.}{2019}]{2019AJ....158...84E}
{Eastwood} M.~W.,  et~al., 2019, \mn@doi [\aj] {10.3847/1538-3881/ab2629},
  \href {https://ui.adsabs.harvard.edu/abs/2019AJ....158...84E} {158, 84}

\bibitem[\protect\citeauthoryear{{Ewall-Wice}, {Chang}, {Lazio}, {Dor{\'e}},
  {Seiffert}  \& {Monsalve}}{{Ewall-Wice} et~al.}{2018}]{2018ApJ...868...63E}
{Ewall-Wice} A.,  {Chang} T.~C.,  {Lazio} J.,  {Dor{\'e}} O.,  {Seiffert} M.,
  {Monsalve} R.~A.,  2018, \mn@doi [\apj] {10.3847/1538-4357/aae51d}, \href
  {https://ui.adsabs.harvard.edu/abs/2018ApJ...868...63E} {868, 63}

\bibitem[\protect\citeauthoryear{{Fan} et~al.,}{{Fan} et~al.}{2006}]{Fan06b}
{Fan} X.,  et~al., 2006, \mn@doi [\aj] {10.1086/504836}, \href
  {http://adsabs.harvard.edu/abs/2006AJ....132..117F} {132, 117}

\bibitem[\protect\citeauthoryear{{Feng} \& {Holder}}{{Feng} \&
  {Holder}}{2018}]{2018ApJ...858L..17F}
{Feng} C.,  {Holder} G.,  2018, \mn@doi [\apjl] {10.3847/2041-8213/aac0fe},
  \href {http://adsabs.harvard.edu/abs/2018ApJ...858L..17F} {858, L17}

\bibitem[\protect\citeauthoryear{{Fialkov} \& {Barkana}}{{Fialkov} \&
  {Barkana}}{2019}]{2019MNRAS.486.1763F}
{Fialkov} A.,  {Barkana} R.,  2019, \mn@doi [\mnras] {10.1093/mnras/stz873},
  \href {https://ui.adsabs.harvard.edu/abs/2019MNRAS.486.1763F} {486, 1763}

\bibitem[\protect\citeauthoryear{{Fialkov}, {Barkana}  \& {Cohen}}{{Fialkov}
  et~al.}{2018}]{2018PhRvL.121a1101F}
{Fialkov} A.,  {Barkana} R.,   {Cohen} A.,  2018, \mn@doi [Physical Review
  Letters] {10.1103/PhysRevLett.121.011101}, \href
  {http://adsabs.harvard.edu/abs/2018PhRvL.121a1101F} {121, 011101}

\bibitem[\protect\citeauthoryear{{Fioc} \& {Rocca-Volmerange}}{{Fioc} \&
  {Rocca-Volmerange}}{1997}]{Fioc97}
{Fioc} M.,  {Rocca-Volmerange} B.,  1997, \aap, \href
  {http://adsabs.harvard.edu/abs/1997A\%26A...326..950F} {326, 950}

\bibitem[\protect\citeauthoryear{{Fixsen} et~al.,}{{Fixsen}
  et~al.}{2011}]{2011ApJ...734....5F}
{Fixsen} D.~J.,  et~al., 2011, \mn@doi [\apj] {10.1088/0004-637X/734/1/5},
  \href {https://ui.adsabs.harvard.edu/abs/2011ApJ...734....5F} {734, 5}

\bibitem[\protect\citeauthoryear{{Foreman-Mackey}, {Hogg}, {Lang}  \&
  {Goodman}}{{Foreman-Mackey} et~al.}{2013}]{emcee2013paper}
{Foreman-Mackey} D.,  {Hogg} D.~W.,  {Lang} D.,   {Goodman} J.,  2013, \mn@doi
  [\pasp] {10.1086/670067}, \href
  {https://ui.adsabs.harvard.edu/abs/2013PASP..125..306F} {125, 306}

\bibitem[\protect\citeauthoryear{{Furlanetto}, {Oh}  \& {Briggs}}{{Furlanetto}
  et~al.}{2006}]{Furlanetto2006}
{Furlanetto} S.~R.,  {Oh} S.~P.,   {Briggs} F.~H.,  2006, \mn@doi [\physrep]
  {10.1016/j.physrep.2006.08.002}, \href
  {http://adsabs.harvard.edu/abs/2006PhR...433..181F} {433, 181}

\bibitem[\protect\citeauthoryear{{Gehlot} et~al.,}{{Gehlot}
  et~al.}{2019}]{2019MNRAS.488.4271G}
{Gehlot} B.~K.,  et~al., 2019, \mn@doi [\mnras] {10.1093/mnras/stz1937}, \href
  {https://ui.adsabs.harvard.edu/abs/2019MNRAS.488.4271G} {488, 4271}

\bibitem[\protect\citeauthoryear{{Gehlot} et~al.,}{{Gehlot}
  et~al.}{2020}]{2020MNRAS.499.4158G}
{Gehlot} B.~K.,  et~al., 2020, \mn@doi [\mnras] {10.1093/mnras/staa3093}, \href
  {https://ui.adsabs.harvard.edu/abs/2020MNRAS.499.4158G} {499, 4158}

\bibitem[\protect\citeauthoryear{{Ghara} \& {Choudhury}}{{Ghara} \&
  {Choudhury}}{2020}]{2020MNRAS.496..739G}
{Ghara} R.,  {Choudhury} T.~R.,  2020, \mn@doi [\mnras]
  {10.1093/mnras/staa1599}, \href
  {https://ui.adsabs.harvard.edu/abs/2020MNRAS.496..739G} {496, 739}

\bibitem[\protect\citeauthoryear{{Ghara} \& {Mellema}}{{Ghara} \&
  {Mellema}}{2020}]{2020MNRAS.492..634G}
{Ghara} R.,  {Mellema} G.,  2020, \mn@doi [\mnras] {10.1093/mnras/stz3513},
  \href {https://ui.adsabs.harvard.edu/abs/2020MNRAS.492..634G} {492, 634}

\bibitem[\protect\citeauthoryear{{Ghara}, {Choudhury}  \& {Datta}}{{Ghara}
  et~al.}{2015a}]{ghara15a}
{Ghara} R.,  {Choudhury} T.~R.,   {Datta} K.~K.,  2015a, \mn@doi [\mnras]
  {10.1093/mnras/stu2512}, \href
  {http://adsabs.harvard.edu/abs/2015MNRAS.447.1806G} {447, 1806}

\bibitem[\protect\citeauthoryear{{Ghara}, {Datta}  \& {Choudhury}}{{Ghara}
  et~al.}{2015b}]{ghara15b}
{Ghara} R.,  {Datta} K.~K.,   {Choudhury} T.~R.,  2015b, \mn@doi [\mnras]
  {10.1093/mnras/stv1855}, \href
  {http://adsabs.harvard.edu/abs/2015MNRAS.453.3143G} {453, 3143}

\bibitem[\protect\citeauthoryear{{Ghara}, {Choudhury}  \& {Datta}}{{Ghara}
  et~al.}{2016}]{ghara15c}
{Ghara} R.,  {Choudhury} T.~R.,   {Datta} K.~K.,  2016, \mn@doi [\mnras]
  {10.1093/mnras/stw953}, \href
  {http://adsabs.harvard.edu/abs/2016MNRAS.460..827G} {460, 827}

\bibitem[\protect\citeauthoryear{{Ghara}, {Choudhury}, {Datta}  \&
  {Choudhuri}}{{Ghara} et~al.}{2017}]{ghara16}
{Ghara} R.,  {Choudhury} T.~R.,  {Datta} K.~K.,   {Choudhuri} S.,  2017,
  \mn@doi [\mnras] {10.1093/mnras/stw2494}, \href
  {http://adsabs.harvard.edu/abs/2017MNRAS.464.2234G} {464, 2234}

\bibitem[\protect\citeauthoryear{{Ghara}, {Mellema}, {Giri}, {Choudhury},
  {Datta}  \& {Majumdar}}{{Ghara} et~al.}{2018}]{ghara18}
{Ghara} R.,  {Mellema} G.,  {Giri} S.~K.,  {Choudhury} T.~R.,  {Datta} K.~K.,
  {Majumdar} S.,  2018, \mn@doi [\mnras] {10.1093/mnras/sty314}, \href
  {http://adsabs.harvard.edu/abs/2018MNRAS.476.1741G} {476, 1741}

\bibitem[\protect\citeauthoryear{{Ghara} et~al.,}{{Ghara}
  et~al.}{2020}]{2020MNRAS.493.4728G}
{Ghara} R.,  et~al., 2020, \mn@doi [\mnras] {10.1093/mnras/staa487}, \href
  {https://ui.adsabs.harvard.edu/abs/2020MNRAS.493.4728G} {493, 4728}

\bibitem[\protect\citeauthoryear{Giri}{Giri}{}]{Giri1289039}
Giri S.~K., , PhD thesis, Stockholm University, Department of Astronomy, 2019.
  , p. 68, ISBN 978-91-7797-611-0

\bibitem[\protect\citeauthoryear{{Giri} \& {Mellema}}{{Giri} \&
  {Mellema}}{2020}]{giri2020BettiNumbers}
{Giri} S.~K.,  {Mellema} G.,  2020, arXiv e-prints, \href
  {https://ui.adsabs.harvard.edu/abs/2020arXiv201212908G} {p. arXiv:2012.12908}

\bibitem[\protect\citeauthoryear{{Giri}, {Mellema}, {Dixon}  \& {Iliev}}{{Giri}
  et~al.}{2018a}]{giri2017bubble}
{Giri} S.~K.,  {Mellema} G.,  {Dixon} K.~L.,   {Iliev} I.~T.,  2018a, \mn@doi
  [\mnras] {10.1093/mnras/stx2539}, \href
  {http://adsabs.harvard.edu/abs/2018MNRAS.473.2949G} {473, 2949}

\bibitem[\protect\citeauthoryear{{Giri}, {Mellema}  \& {Ghara}}{{Giri}
  et~al.}{2018b}]{2018arXiv180106550G}
{Giri} S.~K.,  {Mellema} G.,   {Ghara} R.,  2018b, \mn@doi [\mnras]
  {10.1093/mnras/sty1786}, \href
  {http://adsabs.harvard.edu/abs/2018MNRAS.479.5596G} {479, 5596}

\bibitem[\protect\citeauthoryear{{Giri}, {Mellema}, {Aldheimer}, {Dixon}  \&
  {Iliev}}{{Giri} et~al.}{2019a}]{2019MNRAS.489.1590G}
{Giri} S.~K.,  {Mellema} G.,  {Aldheimer} T.,  {Dixon} K.~L.,   {Iliev} I.~T.,
  2019a, \mn@doi [\mnras] {10.1093/mnras/stz2224}, \href
  {https://ui.adsabs.harvard.edu/abs/2019MNRAS.489.1590G} {489, 1590}

\bibitem[\protect\citeauthoryear{{Giri}, {D'Aloisio}, {Mellema}, {Komatsu},
  {Ghara}  \& {Majumdar}}{{Giri} et~al.}{2019b}]{2019JCAP...02..058G}
{Giri} S.~K.,  {D'Aloisio} A.,  {Mellema} G.,  {Komatsu} E.,  {Ghara} R.,
  {Majumdar} S.,  2019b, \mn@doi [Journal of Cosmology and Astro-Particle
  Physics] {10.1088/1475-7516/2019/02/058}, \href
  {https://ui.adsabs.harvard.edu/abs/2019JCAP...02..058G} {2019, 058}

\bibitem[\protect\citeauthoryear{Greig \& Mesinger}{Greig \&
  Mesinger}{2015}]{Greig201521CMMC:Signal}
Greig B.,  Mesinger A.,  2015, \mn@doi [Monthly Notices of the Royal
  Astronomical Society] {10.1093/mnras/stv571}, 449, 4246

\bibitem[\protect\citeauthoryear{{Greig} \& {Mesinger}}{{Greig} \&
  {Mesinger}}{2017}]{2017MNRAS.472.2651G}
{Greig} B.,  {Mesinger} A.,  2017, \mn@doi [\mnras] {10.1093/mnras/stx2118},
  \href {https://ui.adsabs.harvard.edu/abs/2017MNRAS.472.2651G} {472, 2651}

\bibitem[\protect\citeauthoryear{{Greig}, {Mesinger}, {Haiman}  \&
  {Simcoe}}{{Greig} et~al.}{2017}]{2017MNRAS.466.4239G}
{Greig} B.,  {Mesinger} A.,  {Haiman} Z.,   {Simcoe} R.~A.,  2017, \mn@doi
  [\mnras] {10.1093/mnras/stw3351}, \href
  {https://ui.adsabs.harvard.edu/abs/2017MNRAS.466.4239G} {466, 4239}

\bibitem[\protect\citeauthoryear{{Greig}, {Mesinger}  \& {Ba{\~n}ados}}{{Greig}
  et~al.}{2019}]{2019MNRAS.484.5094G}
{Greig} B.,  {Mesinger} A.,   {Ba{\~n}ados} E.,  2019, \mn@doi [\mnras]
  {10.1093/mnras/stz230}, \href
  {https://ui.adsabs.harvard.edu/abs/2019MNRAS.484.5094G} {484, 5094}

\bibitem[\protect\citeauthoryear{{Greig}, {Trott}, {Barry}, {Mutch}, {Pindor},
  {Webster}  \& {Wyithe}}{{Greig} et~al.}{2021a}]{2020arXiv200802639G}
{Greig} B.,  {Trott} C.~M.,  {Barry} N.,  {Mutch} S.~J.,  {Pindor} B.,
  {Webster} R.~L.,   {Wyithe} J. S.~B.,  2021a, \mn@doi [\mnras]
  {10.1093/mnras/staa3494}, \href
  {https://ui.adsabs.harvard.edu/abs/2021MNRAS.500.5322G} {500, 5322}

\bibitem[\protect\citeauthoryear{{Greig} et~al.,}{{Greig}
  et~al.}{2021b}]{2020arXiv200603203G}
{Greig} B.,  et~al., 2021b, \mn@doi [\mnras] {10.1093/mnras/staa3593}, \href
  {https://ui.adsabs.harvard.edu/abs/2021MNRAS.501....1G} {501, 1}

\bibitem[\protect\citeauthoryear{{Gupta} et~al.,}{{Gupta}
  et~al.}{2017}]{2017CSci..113..707G}
{Gupta} Y.,  et~al., 2017, Current Science, \href
  {https://ui.adsabs.harvard.edu/abs/2017CSci..113..707G} {113, 707}

\bibitem[\protect\citeauthoryear{{Harker} et~al.,}{{Harker}
  et~al.}{2009}]{2009MNRAS.397.1138H}
{Harker} G.,  et~al., 2009, \mn@doi [\mnras]
  {10.1111/j.1365-2966.2009.15081.x}, \href
  {https://ui.adsabs.harvard.edu/abs/2009MNRAS.397.1138H} {397, 1138}

\bibitem[\protect\citeauthoryear{{Hasegawa} \& {Semelin}}{{Hasegawa} \&
  {Semelin}}{2013}]{2013MNRAS.428..154H}
{Hasegawa} K.,  {Semelin} B.,  2013, \mn@doi [\mnras] {10.1093/mnras/sts021},
  \href {https://ui.adsabs.harvard.edu/abs/2013MNRAS.428..154H} {428, 154}

\bibitem[\protect\citeauthoryear{{Hills}, {Kulkarni}, {Meerburg}  \&
  {Puchwein}}{{Hills} et~al.}{2018}]{2018Natur.564E..32H}
{Hills} R.,  {Kulkarni} G.,  {Meerburg} P.~D.,   {Puchwein} E.,  2018, \mn@doi
  [\nat] {10.1038/s41586-018-0796-5}, \href
  {http://adsabs.harvard.edu/abs/2018Natur.564E..32H} {564, E32}

\bibitem[\protect\citeauthoryear{{Hinshaw} et~al.,}{{Hinshaw}
  et~al.}{2013}]{2013ApJS..208...19H}
{Hinshaw} G.,  et~al., 2013, \mn@doi [\apjs] {10.1088/0067-0049/208/2/19},
  \href {http://adsabs.harvard.edu/abs/2013ApJS..208...19H} {208, 19}

\bibitem[\protect\citeauthoryear{{Hothi} et~al.,}{{Hothi}
  et~al.}{2021}]{2021MNRAS.500.2264H}
{Hothi} I.,  et~al., 2021, \mn@doi [\mnras] {10.1093/mnras/staa3446}, \href
  {https://ui.adsabs.harvard.edu/abs/2021MNRAS.500.2264H} {500, 2264}

\bibitem[\protect\citeauthoryear{{Islam}, {Ghara}, {Paul}, {Choudhury}  \&
  {Nath}}{{Islam} et~al.}{2019}]{2019MNRAS.487.2785I}
{Islam} N.,  {Ghara} R.,  {Paul} B.,  {Choudhury} T.~R.,   {Nath} B.~B.,  2019,
  \mn@doi [\mnras] {10.1093/mnras/stz1446}, \href
  {https://ui.adsabs.harvard.edu/abs/2019MNRAS.487.2785I} {487, 2785}

\bibitem[\protect\citeauthoryear{{Jensen} et~al.,}{{Jensen}
  et~al.}{2013}]{Jensen13}
{Jensen} H.,  et~al., 2013, \mn@doi [\mnras] {10.1093/mnras/stt1341}, \href
  {http://adsabs.harvard.edu/abs/2013MNRAS.435..460J} {435, 460}

\bibitem[\protect\citeauthoryear{{Kamran}, {Ghara}, {Majumdar}, {Mondal},
  {Mellema}, {Bharadwaj}, {Pritchard}  \& {Iliev}}{{Kamran}
  et~al.}{2021}]{2021MNRAS.502.3800K}
{Kamran} M.,  {Ghara} R.,  {Majumdar} S.,  {Mondal} R.,  {Mellema} G.,
  {Bharadwaj} S.,  {Pritchard} J.~R.,   {Iliev} I.~T.,  2021, \mn@doi [\mnras]
  {10.1093/mnras/stab216}, \href
  {https://ui.adsabs.harvard.edu/abs/2021MNRAS.502.3800K} {502, 3800}

\bibitem[\protect\citeauthoryear{{Kapahtia}, {Chingangbam}, {Ghara}, {Appleby}
  \& {Choudhury}}{{Kapahtia} et~al.}{2021}]{2021arXiv210103962K}
{Kapahtia} A.,  {Chingangbam} P.,  {Ghara} R.,  {Appleby} S.,   {Choudhury}
  T.~R.,  2021, arXiv e-prints, \href
  {https://ui.adsabs.harvard.edu/abs/2021arXiv210103962K} {p. arXiv:2101.03962}

\bibitem[\protect\citeauthoryear{{Kern}, {Parsons}, {Dillon}, {Lanman},
  {Fagnoni}  \& {de Lera Acedo}}{{Kern} et~al.}{2019}]{2019ApJ...884..105K}
{Kern} N.~S.,  {Parsons} A.~R.,  {Dillon} J.~S.,  {Lanman} A.~E.,  {Fagnoni}
  N.,   {de Lera Acedo} E.,  2019, \mn@doi [\apj] {10.3847/1538-4357/ab3e73},
  \href {https://ui.adsabs.harvard.edu/abs/2019ApJ...884..105K} {884, 105}

\bibitem[\protect\citeauthoryear{{Kolopanis} et~al.,}{{Kolopanis}
  et~al.}{2019}]{2019ApJ...883..133K}
{Kolopanis} M.,  et~al., 2019, \mn@doi [\apj] {10.3847/1538-4357/ab3e3a}, \href
  {https://ui.adsabs.harvard.edu/abs/2019ApJ...883..133K} {883, 133}

\bibitem[\protect\citeauthoryear{{Krause}, {Thomas}, {Zaroubi}  \&
  {Abdalla}}{{Krause} et~al.}{2018}]{2018NewA...64....9K}
{Krause} F.,  {Thomas} R.~M.,  {Zaroubi} S.,   {Abdalla} F.~B.,  2018, \mn@doi
  [\na] {10.1016/j.newast.2018.03.004}, \href
  {https://ui.adsabs.harvard.edu/abs/2018NewA...64....9K} {64, 9}

\bibitem[\protect\citeauthoryear{{Liu}, {Parsons}  \& {Trott}}{{Liu}
  et~al.}{2014}]{2014PhRvD..90b3019L}
{Liu} A.,  {Parsons} A.~R.,   {Trott} C.~M.,  2014, \mn@doi [\prd]
  {10.1103/PhysRevD.90.023019}, \href
  {https://ui.adsabs.harvard.edu/abs/2014PhRvD..90b3019L} {90, 023019}

\bibitem[\protect\citeauthoryear{{Madau}, {Meiksin}  \& {Rees}}{{Madau}
  et~al.}{1997}]{madau1997}
{Madau} P.,  {Meiksin} A.,   {Rees} M.~J.,  1997, \apj, \href
  {http://adsabs.harvard.edu/abs/1997ApJ...475..429M} {475, 429}

\bibitem[\protect\citeauthoryear{{Majumdar}, {Bharadwaj}  \&
  {Choudhury}}{{Majumdar} et~al.}{2012}]{2012MNRAS.426.3178M}
{Majumdar} S.,  {Bharadwaj} S.,   {Choudhury} T.~R.,  2012, \mn@doi [\mnras]
  {10.1111/j.1365-2966.2012.21914.x}, \href
  {http://adsabs.harvard.edu/abs/2012MNRAS.426.3178M} {426, 3178}

\bibitem[\protect\citeauthoryear{{Majumdar}, {Datta}, {Ghara}, {Mondal},
  {Choudhury}, {Bharadwaj}, {Ali}  \& {Datta}}{{Majumdar}
  et~al.}{2016}]{2016JApA...37...32M}
{Majumdar} S.,  {Datta} K.~K.,  {Ghara} R.,  {Mondal} R.,  {Choudhury} T.~R.,
  {Bharadwaj} S.,  {Ali} S.~S.,   {Datta} A.,  2016, \mn@doi [Journal of
  Astrophysics and Astronomy] {10.1007/s12036-016-9402-0}, \href
  {https://ui.adsabs.harvard.edu/abs/2016JApA...37...32M} {37, 32}

\bibitem[\protect\citeauthoryear{{McGreer}, {Mesinger}  \&
  {D'Odorico}}{{McGreer} et~al.}{2015}]{2015MNRAS.447..499M}
{McGreer} I.~D.,  {Mesinger} A.,   {D'Odorico} V.,  2015, \mn@doi [\mnras]
  {10.1093/mnras/stu2449}, \href
  {https://ui.adsabs.harvard.edu/abs/2015MNRAS.447..499M} {447, 499}

\bibitem[\protect\citeauthoryear{Mebane, Mirocha  \& Furlanetto}{Mebane
  et~al.}{2020}]{Mebane2020}
Mebane R.~H.,  Mirocha J.,   Furlanetto S.~R.,  2020, \mn@doi [Monthly Notices
  of the Royal Astronomical Society] {10.1093/mnras/staa280}, 493, 1217

\bibitem[\protect\citeauthoryear{{Mellema}, {Koopmans}, {Shukla}, {Datta},
  {Mesinger}  \& {Majumdar}}{{Mellema} et~al.}{2015}]{2015aska.confE..10M}
{Mellema} G.,  {Koopmans} L.,  {Shukla} H.,  {Datta} K.~K.,  {Mesinger} A.,
  {Majumdar} S.,  2015, Advancing Astrophysics with the Square Kilometre Array
  (AASKA14), \href {http://adsabs.harvard.edu/abs/2015aska.confE..10M} {p.~10}

\bibitem[\protect\citeauthoryear{{Mertens}, {Ghosh}  \& {Koopmans}}{{Mertens}
  et~al.}{2018}]{2018MNRAS.478.3640M}
{Mertens} F.~G.,  {Ghosh} A.,   {Koopmans} L.~V.~E.,  2018, \mn@doi [\mnras]
  {10.1093/mnras/sty1207}, \href
  {https://ui.adsabs.harvard.edu/abs/2018MNRAS.478.3640M} {478, 3640}

\bibitem[\protect\citeauthoryear{{Mertens} et~al.,}{{Mertens}
  et~al.}{2020}]{2020MNRAS.493.1662M}
{Mertens} F.~G.,  et~al., 2020, \mn@doi [\mnras] {10.1093/mnras/staa327}, \href
  {https://ui.adsabs.harvard.edu/abs/2020MNRAS.493.1662M} {493, 1662}

\bibitem[\protect\citeauthoryear{{Mesinger} \& {Furlanetto}}{{Mesinger} \&
  {Furlanetto}}{2007}]{2007ApJ...669..663M}
{Mesinger} A.,  {Furlanetto} S.,  2007, \mn@doi [\apj] {10.1086/521806}, \href
  {http://cdsads.u-strasbg.fr/abs/2007ApJ...669..663M} {669, 663}

\bibitem[\protect\citeauthoryear{{Mevius} et~al.,}{{Mevius}
  et~al.}{2016}]{2016RaSc...51..927M}
{Mevius} M.,  et~al., 2016, \mn@doi [Radio Science] {10.1002/2016RS006028},
  \href {https://ui.adsabs.harvard.edu/abs/2016RaSc...51..927M} {51, 927}

\bibitem[\protect\citeauthoryear{{Mineo}, {Gilfanov}  \& {Sunyaev}}{{Mineo}
  et~al.}{2012}]{2012MNRAS.419.2095M}
{Mineo} S.,  {Gilfanov} M.,   {Sunyaev} R.,  2012, \mn@doi [\mnras]
  {10.1111/j.1365-2966.2011.19862.x}, \href
  {https://ui.adsabs.harvard.edu/abs/2012MNRAS.419.2095M} {419, 2095}

\bibitem[\protect\citeauthoryear{{Mitra}, {Choudhury}  \& {Ferrara}}{{Mitra}
  et~al.}{2015}]{Mitra15}
{Mitra} S.,  {Choudhury} T.~R.,   {Ferrara} A.,  2015, \mn@doi [\mnras]
  {10.1093/mnrasl/slv134}, \href
  {http://adsabs.harvard.edu/abs/2015MNRAS.454L..76M} {454, L76}

\bibitem[\protect\citeauthoryear{{Mondal} et~al.,}{{Mondal}
  et~al.}{2020}]{2020MNRAS.498.4178M}
{Mondal} R.,  et~al., 2020, \mn@doi [\mnras] {10.1093/mnras/staa2422}, \href
  {https://ui.adsabs.harvard.edu/abs/2020MNRAS.498.4178M} {498, 4178}

\bibitem[\protect\citeauthoryear{{Monsalve}, {Rogers}, {Bowman}  \&
  {Mozdzen}}{{Monsalve} et~al.}{2017}]{monsalve2017}
{Monsalve} R.~A.,  {Rogers} A. E.~E.,  {Bowman} J.~D.,   {Mozdzen} T.~J.,
  2017, \mn@doi [\apj] {10.3847/1538-4357/835/1/49}, \href
  {https://ui.adsabs.harvard.edu/\#abs/2017ApJ...835...49M} {835, 49}

\bibitem[\protect\citeauthoryear{{Monsalve}, {Greig}, {Bowman}, {Mesinger},
  {Rogers}, {Mozdzen}, {Kern}  \& {Mahesh}}{{Monsalve}
  et~al.}{2018}]{2018ApJ...863...11M}
{Monsalve} R.~A.,  {Greig} B.,  {Bowman} J.~D.,  {Mesinger} A.,  {Rogers} A.
  E.~E.,  {Mozdzen} T.~J.,  {Kern} N.~S.,   {Mahesh} N.,  2018, \mn@doi [\apj]
  {10.3847/1538-4357/aace54}, \href
  {https://ui.adsabs.harvard.edu/abs/2018ApJ...863...11M} {863, 11}

\bibitem[\protect\citeauthoryear{{Mu{\~n}oz} \& {Loeb}}{{Mu{\~n}oz} \&
  {Loeb}}{2018}]{2018arXiv180210094M}
{Mu{\~n}oz} J.~B.,  {Loeb} A.,  2018, arXiv e-prints, \href
  {https://ui.adsabs.harvard.edu/\#abs/2018arXiv180210094M} {p.
  arXiv:1802.10094}

\bibitem[\protect\citeauthoryear{{Nambissan} et~al.,}{{Nambissan}
  et~al.}{2021}]{saras3}
{Nambissan} J.,  et~al., 2021, \mn@doi [Experimental Astronomy]
  {10.1007/s10686-020-09697-2}, 1572-9508

\bibitem[\protect\citeauthoryear{{Ouchi} et~al.,}{{Ouchi}
  et~al.}{2010}]{2010ApJ...723..869O}
{Ouchi} M.,  et~al., 2010, \mn@doi [\apj] {10.1088/0004-637X/723/1/869}, \href
  {https://ui.adsabs.harvard.edu/abs/2010ApJ...723..869O} {723, 869}

\bibitem[\protect\citeauthoryear{{Paciga} et~al.,}{{Paciga}
  et~al.}{2013}]{paciga13}
{Paciga} G.,  et~al., 2013, \mn@doi [\mnras] {10.1093/mnras/stt753}, \href
  {http://adsabs.harvard.edu/abs/2013MNRAS.433..639P} {433, 639}

\bibitem[\protect\citeauthoryear{Park, Mesinger, Greig  \& Gillet}{Park
  et~al.}{2019}]{Park2019InferringSignal}
Park J.,  Mesinger A.,  Greig B.,   Gillet N.,  2019, \mn@doi [Monthly Notices
  of the Royal Astronomical Society] {10.1093/mnras/stz032}, 484, 933

\bibitem[\protect\citeauthoryear{{Parsons} et~al.,}{{Parsons}
  et~al.}{2014}]{parsons13}
{Parsons} A.~R.,  et~al., 2014, \mn@doi [\apj] {10.1088/0004-637X/788/2/106},
  \href {http://adsabs.harvard.edu/abs/2014ApJ...788..106P} {788, 106}

\bibitem[\protect\citeauthoryear{{Patil} et~al.,}{{Patil}
  et~al.}{2017}]{2017ApJ...838...65P}
{Patil} A.~H.,  et~al., 2017, \mn@doi [\apj] {10.3847/1538-4357/aa63e7}, \href
  {http://adsabs.harvard.edu/abs/2017ApJ...838...65P} {838, 65}

\bibitem[\protect\citeauthoryear{{Patra}, {Subrahmanyan}, {Sethi}, {Udaya
  Shankar}  \& {Raghunathan}}{{Patra} et~al.}{2015}]{2015ApJ...801..138P}
{Patra} N.,  {Subrahmanyan} R.,  {Sethi} S.,  {Udaya Shankar} N.,
  {Raghunathan} A.,  2015, \mn@doi [\apj] {10.1088/0004-637X/801/2/138}, \href
  {http://adsabs.harvard.edu/abs/2015ApJ...801..138P} {801, 138}

\bibitem[\protect\citeauthoryear{{Philip} et~al.,}{{Philip}
  et~al.}{2019}]{2019JAI.....850004P}
{Philip} L.,  et~al., 2019, \mn@doi [Journal of Astronomical Instrumentation]
  {10.1142/S2251171719500041}, \href
  {https://ui.adsabs.harvard.edu/abs/2019JAI.....850004P} {8, 1950004}

\bibitem[\protect\citeauthoryear{{Planck Collaboration} et~al.,}{{Planck
  Collaboration} et~al.}{2020}]{2020A&A...641A...6P}
{Planck Collaboration} et~al., 2020, \mn@doi [\aap]
  {10.1051/0004-6361/201833910}, \href
  {https://ui.adsabs.harvard.edu/abs/2020A&A...641A...6P} {641, A6}

\bibitem[\protect\citeauthoryear{{Price} et~al.,}{{Price}
  et~al.}{2018}]{price2018}
{Price} D.~C.,  et~al., 2018, \mn@doi [\mnras] {10.1093/mnras/sty1244}, \href
  {https://ui.adsabs.harvard.edu/\#abs/2018MNRAS.478.4193P} {478, 4193}

\bibitem[\protect\citeauthoryear{{Pritchard} \& {Loeb}}{{Pritchard} \&
  {Loeb}}{2012}]{Pritchard12}
{Pritchard} J.~R.,  {Loeb} A.,  2012, \mn@doi [Reports on Progress in Physics]
  {10.1088/0034-4885/75/8/086901}, \href
  {http://adsabs.harvard.edu/abs/2012RPPh...75h6901P} {75, 086901}

\bibitem[\protect\citeauthoryear{{Rasmussen} \& {Williams}}{{Rasmussen} \&
  {Williams}}{2006}]{GPR2006}
{Rasmussen} C.~E.,  {Williams} C.~K.~I.,  2006, {Gaussian Processes for Machine
  Learning}.
The MIT Press, 2006. ISBN 0-262-18253-X

\bibitem[\protect\citeauthoryear{{Reis}, {Fialkov}  \& {Barkana}}{{Reis}
  et~al.}{2020a}]{2020MNRAS.499.5993R}
{Reis} I.,  {Fialkov} A.,   {Barkana} R.,  2020a, \mn@doi [\mnras]
  {10.1093/mnras/staa3091}, \href
  {https://ui.adsabs.harvard.edu/abs/2020MNRAS.499.5993R} {499, 5993}

\bibitem[\protect\citeauthoryear{{Reis}, {Fialkov}  \& {Barkana}}{{Reis}
  et~al.}{2020b}]{Reis2020Highredshift}
{Reis} I.,  {Fialkov} A.,   {Barkana} R.,  2020b, \mn@doi [\mnras]
  {10.1093/mnras/staa3091}, \href
  {https://ui.adsabs.harvard.edu/abs/2020MNRAS.499.5993R} {499, 5993}

\bibitem[\protect\citeauthoryear{{Ross}, {Dixon}, {Ghara}, {Iliev}  \&
  {Mellema}}{{Ross} et~al.}{2019}]{2019MNRAS.tmp.1183R}
{Ross} H.~E.,  {Dixon} K.~L.,  {Ghara} R.,  {Iliev} I.~T.,   {Mellema} G.,
  2019, \mn@doi [\mnras] {10.1093/mnras/stz1220}, \href
  {https://ui.adsabs.harvard.edu/abs/2019MNRAS.tmp.1183R} {p.~1183}

\bibitem[\protect\citeauthoryear{{Ross}, {Giri}, {Dixon}, {Ghara}, {Iliev}  \&
  {Mellema}}{{Ross} et~al.}{2020}]{2020arXiv201103558R}
{Ross} H.~E.,  {Giri} S.~K.,  {Dixon} K.~L.,  {Ghara} R.,  {Iliev} I.~T.,
  {Mellema} G.,  2020, arXiv e-prints, \href
  {https://ui.adsabs.harvard.edu/abs/2020arXiv201103558R} {p. arXiv:2011.03558}

\bibitem[\protect\citeauthoryear{{Schneider}, {Giri}  \& {Mirocha}}{{Schneider}
  et~al.}{2020}]{2020arXiv201112308S}
{Schneider} A.,  {Giri} S.,   {Mirocha} J.,  2020, arXiv e-prints, \href
  {https://ui.adsabs.harvard.edu/abs/2020arXiv201112308S} {p. arXiv:2011.12308}

\bibitem[\protect\citeauthoryear{{Singh} \& {Subrahmanyan}}{{Singh} \&
  {Subrahmanyan}}{2019}]{2019ApJ...880...26S}
{Singh} S.,  {Subrahmanyan} R.,  2019, \mn@doi [\apj]
  {10.3847/1538-4357/ab2879}, \href
  {https://ui.adsabs.harvard.edu/abs/2019ApJ...880...26S} {880, 26}

\bibitem[\protect\citeauthoryear{{Singh} et~al.,}{{Singh}
  et~al.}{2017}]{singh2017}
{Singh} S.,  et~al., 2017, \mn@doi [\apj] {10.3847/2041-8213/aa831b}, \href
  {https://ui.adsabs.harvard.edu/\#abs/2017ApJ...845L..12S} {845, L12}

\bibitem[\protect\citeauthoryear{{Sokolowski} et~al.,}{{Sokolowski}
  et~al.}{2015}]{2015PASA...32....4S}
{Sokolowski} M.,  et~al., 2015, \mn@doi [\pasa] {10.1017/pasa.2015.3}, \href
  {http://adsabs.harvard.edu/abs/2015PASA...32....4S} {32, e004}

\bibitem[\protect\citeauthoryear{{Sun} \& {Furlanetto}}{{Sun} \&
  {Furlanetto}}{2016}]{2016MNRAS.460..417S}
{Sun} G.,  {Furlanetto} S.~R.,  2016, \mn@doi [\mnras] {10.1093/mnras/stw980},
  \href {https://ui.adsabs.harvard.edu/abs/2016MNRAS.460..417S} {460, 417}

\bibitem[\protect\citeauthoryear{{Tashiro}, {Kadota}  \& {Silk}}{{Tashiro}
  et~al.}{2014}]{2014PhRvD..90h3522T}
{Tashiro} H.,  {Kadota} K.,   {Silk} J.,  2014, \mn@doi [\prd]
  {10.1103/PhysRevD.90.083522}, \href
  {https://ui.adsabs.harvard.edu/abs/2014PhRvD..90h3522T} {90, 083522}

\bibitem[\protect\citeauthoryear{{Thomas} \& {Zaroubi}}{{Thomas} \&
  {Zaroubi}}{2008}]{Thom08}
{Thomas} R.~M.,  {Zaroubi} S.,  2008, \mn@doi [\mnras]
  {10.1111/j.1365-2966.2007.12767.x}, \href
  {https://ui.adsabs.harvard.edu/abs/2008MNRAS.384.1080T} {384, 1080}

\bibitem[\protect\citeauthoryear{{Thomas} \& {Zaroubi}}{{Thomas} \&
  {Zaroubi}}{2011}]{Thom11}
{Thomas} R.~M.,  {Zaroubi} S.,  2011, \mn@doi [\mnras]
  {10.1111/j.1365-2966.2010.17525.x}, \href
  {https://ui.adsabs.harvard.edu/abs/2011MNRAS.410.1377T} {410, 1377}

\bibitem[\protect\citeauthoryear{{Thomas} et~al.,}{{Thomas}
  et~al.}{2009}]{Thom09}
{Thomas} R.~M.,  et~al., 2009, \mn@doi [\mnras]
  {10.1111/j.1365-2966.2008.14206.x}, \href
  {http://adsabs.harvard.edu/abs/2009MNRAS.393...32T} {393, 32}

\bibitem[\protect\citeauthoryear{{Tingay} et~al.,}{{Tingay}
  et~al.}{2013}]{tingay13}
{Tingay} S.~J.,  et~al., 2013, \mn@doi [Publications of the Astronomical
  Society of Australia (PASA)] {10.1017/pasa.2012.007}, \href
  {http://adsabs.harvard.edu/abs/2013PASA...30....7T} {30, 7}

\bibitem[\protect\citeauthoryear{{Trott} et~al.,}{{Trott}
  et~al.}{2020}]{2020MNRAS.493.4711T}
{Trott} C.~M.,  et~al., 2020, \mn@doi [\mnras] {10.1093/mnras/staa414}, \href
  {https://ui.adsabs.harvard.edu/abs/2020MNRAS.493.4711T} {493, 4711}

\bibitem[\protect\citeauthoryear{{Virtanen} et~al.,}{{Virtanen}
  et~al.}{2020}]{Virtanen2020scipy}
{Virtanen} P.,  et~al., 2020, \mn@doi [Nature Methods]
  {10.1038/s41592-019-0686-2}, \href
  {https://ui.adsabs.harvard.edu/abs/2020NatMe..17..261V} {17, 261}

\bibitem[\protect\citeauthoryear{{Voytek}, {Natarajan}, {J{\'a}uregui
  Garc{\'{\i}}a}, {Peterson}  \& {L{\'o}pez-Cruz}}{{Voytek}
  et~al.}{2014}]{2014ApJ...782L...9V}
{Voytek} T.~C.,  {Natarajan} A.,  {J{\'a}uregui Garc{\'{\i}}a} J.~M.,
  {Peterson} J.~B.,   {L{\'o}pez-Cruz} O.,  2014, \mn@doi [\apjl]
  {10.1088/2041-8205/782/1/L9}, \href
  {http://adsabs.harvard.edu/abs/2014ApJ...782L...9V} {782, L9}

\bibitem[\protect\citeauthoryear{{Wang} et~al.,}{{Wang}
  et~al.}{2020}]{2020ApJ...896...23W}
{Wang} F.,  et~al., 2020, \mn@doi [\apj] {10.3847/1538-4357/ab8c45}, \href
  {https://ui.adsabs.harvard.edu/abs/2020ApJ...896...23W} {896, 23}

\bibitem[\protect\citeauthoryear{Watson, Iliev, D’Aloisio, Knebe, Shapiro  \&
  Yepes}{Watson et~al.}{2013}]{Watson2013TheAges}
Watson W.~A.,  Iliev I.~T.,  D’Aloisio A.,  Knebe A.,  Shapiro P.~R.,   Yepes
  G.,  2013, \mn@doi [Monthly Notices of the Royal Astronomical Society]
  {10.1093/mnras/stt791}, 433, 1230

\bibitem[\protect\citeauthoryear{{Wayth} et~al.,}{{Wayth}
  et~al.}{2018}]{Wayth2018mwa}
{Wayth} R.~B.,  et~al., 2018, \mn@doi [\pasa] {10.1017/pasa.2018.37}, \href
  {https://ui.adsabs.harvard.edu/abs/2018PASA...35...33W} {35, 33}

\bibitem[\protect\citeauthoryear{Weiser \& Zarantonello}{Weiser \&
  Zarantonello}{1988}]{weiser1988note}
Weiser A.,  Zarantonello S.~E.,  1988, Mathematics of Computation, 50, 189

\bibitem[\protect\citeauthoryear{{Wise}, {Demchenko}, {Halicek}, {Norman},
  {Turk}, {Abel}  \& {Smith}}{{Wise} et~al.}{2014}]{2014MNRAS.442.2560W}
{Wise} J.~H.,  {Demchenko} V.~G.,  {Halicek} M.~T.,  {Norman} M.~L.,  {Turk}
  M.~J.,  {Abel} T.,   {Smith} B.~D.,  2014, \mn@doi [\mnras]
  {10.1093/mnras/stu979}, \href
  {http://adsabs.harvard.edu/abs/2014MNRAS.442.2560W} {442, 2560}

\bibitem[\protect\citeauthoryear{{Yang} et~al.,}{{Yang}
  et~al.}{2020}]{2020ApJ...904...26Y}
{Yang} J.,  et~al., 2020, \mn@doi [\apj] {10.3847/1538-4357/abbc1b}, \href
  {https://ui.adsabs.harvard.edu/abs/2020ApJ...904...26Y} {904, 26}

\bibitem[\protect\citeauthoryear{{Zarka}, {Girard}, {Tagger}  \&
  {Denis}}{{Zarka} et~al.}{2012}]{2012sf2a.conf..687Z}
{Zarka} P.,  {Girard} J.~N.,  {Tagger} M.,   {Denis} L.,  2012, in {Boissier}
  S.,  {de Laverny} P.,  {Nardetto} N.,  {Samadi} R.,  {Valls-Gabaud} D.,
  {Wozniak} H.,  eds, SF2A-2012: Proceedings of the Annual meeting of the
  French Society of Astronomy and Astrophysics. pp 687--694

\bibitem[\protect\citeauthoryear{{Zaroubi}}{{Zaroubi}}{2013}]{2013ASSL..396...45Z}
{Zaroubi} S.,  2013, in {Wiklind} T.,  {Mobasher} B.,   {Bromm} V.,  eds,
  Astrophysics and Space Science Library Vol. 396, The First Galaxies. p.~45
  (\mn@eprint {arXiv} {1206.0267}), \mn@doi{10.1007/978-3-642-32362-1_2}

\bibitem[\protect\citeauthoryear{van Haarlem et~al.,}{van Haarlem
  et~al.}{2013}]{vanHaarlem2013LOFAR:ARray}
van Haarlem M.~P.,  et~al., 2013, \mn@doi [Astronomy {\&} Astrophysics]
  {10.1051/0004-6361/201220873}, 556, A2

\makeatother
\end{thebibliography}

\bsp	
\label{lastpage}
\end{document}